\documentclass[aps,prl, twocolumn,showpacs,superscriptaddress,longbibliography]{revtex4-2}
\usepackage{xcolor}
\usepackage{graphicx}
\usepackage{hyperref}
\usepackage{dcolumn}
\usepackage{bm}
\usepackage{amsmath}
\usepackage{bbold}
\usepackage{mathtools}
\usepackage{amsmath, bigstrut}
\usepackage{nicematrix, tikz}
\usetikzlibrary{fit}
\usetikzlibrary{matrix,decorations.pathreplacing}
\usepackage{amssymb}
\usepackage{mathrsfs}
\usepackage{natbib}
\usepackage{notoccite}

\newcommand{\tr}{\text{Tr}}
\newcommand{\nn}{\nonumber \\ }

\newcommand{\im}{\text{Im}}
\newcommand{\re}{\text{Re}}
\usepackage{graphicx}
\usepackage{pstricks}
\usetikzlibrary{arrows.meta, positioning}
\usepackage[normalem]{ulem}
\usepackage{braket}
\usepackage{makecell}

\begin{document}

\preprint{APS/123-QED}

\title{
Quantum-optimal Frequency Estimation of Stochastic AC Fields 
}

\author{Anirban Dey}
\affiliation{School of Mathematical and Physical Sciences, Macquarie University,  NSW 2109, Australia}
\affiliation{ARC Centre of Excellence for Engineered Quantum Systems, Macquarie University,  NSW 2109, Australia}
\author{Sara Mouradian}
\affiliation{Department of Electrical and Computer Engineering, University of Washington - 98195, USA}
\author{Cosmo Lupo}
\affiliation{Dipartimento Interateneo di Fisica, Politecnico di Bari, 70126, Bari, Italy}
\affiliation{Dipartimento Interateneo di Fisica, Università di Bari, 70126, Bari, Italy}
\affiliation{INFN, Sezione di Bari, 70126 Bari, Italy}

\author{Zixin Huang}
\email{zixin.huang@rmit.edu.au}
\affiliation{School of Science, College of STEM, RMIT University, Melbourne, Australia}

\date{\today}

\begin{abstract}

Resolving frequencies in a time-dependent field is classically limited by the measurement bandwidth.
Using tools from quantum metrology and quantum control may overcome this limit, 
yet the full advantage afforded by entanglement so far remains elusive.
Here we map the problem of frequency measurement to that of estimating a global dephasing quantum channel.
In this way, we determine the ultimate quantum limits of {frequency estimation in stochastic AC} sensing.
We find exact {quantum Fisher information bounds} for estimating frequency and frequency differences of stochastic fields. 
In particular, given two close signals with frequency separation $\omega_r$, we find that the quantum Fisher information (QFI) for the separation estimation is approximately $2/\omega_r^2$, {i.e.}~\emph{inversely} proportional to the separation parameter. 
The bounds are achievable in certain regimes by superpositions of Dicke states.
GHZ states are suboptimal but improve precision over unentangled states, achieving Heisenberg scaling in the low-bandwidth limit. 
This work establishes a robust framework for stochastic AC signal sensing that can be extended to arbitrary time-dependent and stochastic fields.
\end{abstract}

\maketitle

Quantum metrology~\cite{giovannetti2006quantum, giovannetti2004quantum, giovannetti2011advances} 
exploits properties that are uniquely quantum -- such as coherence, squeezing and entanglement -- to achieve sensitivity beyond what is classically possible. 
Parameter estimation lies at the heart of a wide range of applications, and the quantum advantage for using spin systems to measure the field amplitudes of DC fields has been extensively studied (see e.g.~Refs.~\cite{RevModPhys.89.035002,degen2008scanning,taylor2008high,PhysRevLett.115.170801,PhysRevA.96.042319,bonato2016optimized,jones2009magnetic}).
%
%
%
%
{In contrast, the
estimation of frequencies—despite its importance to noise metrology for quantum computing, precision GPS, and quantum control \cite{riberi2025optimal, gefen2019, Bonizzoni2024, WangPRX,Iemini2024, WOLF2021}—has received less attention. Of particular interest is \textit{global dephasing}, which is the main noise mechanism that afflicts quantum information systems such as Bose-Einstein condensates \cite{khodorkovsky2009decoherence,liu2010quantum}, trapped ions, and cold neutral atoms \cite{carnio2015robust,dorner2012quantum,gross2010nonlinear} .}

In this work, we address the problem of estimating the frequency components of time-varying, potentially multi-frequency AC signals.
 In classical systems, closely spaced frequencies are difficult to resolve due to bandwidth limitations, typically expressed as~\cite{RevModPhys.89.035002}:
\begin{align}
\Delta \omega \approx 1/t \, , 
\end{align}
where $\Delta \omega$ is the frequency uncertainty and $t$ is the evolution time. Resolving two frequencies then requires an observation time {inversely proportional to their frequency separation}. {In this work, we reduce the problem of frequency estimation to that of estimating noise parameters in a dephasing channel.}


\begin{table*}[t!]
\begin{center}
\renewcommand{\arraystretch}{1.7}
\begin{tabular}{| c | c c c c | c | }
\hline
    & coherent (qubit) & {stochastic} (qubit)  & \makecell {coherent\\ (N-qubit GHZ)} & \makecell{{stochastic}\\ (N-qubit GHZ)} & QFI bound (incoherent) \\ 
    \hline
 single-frequency   &    $B^2 t^4$\cite{pang2017optimal}    
                    &    $\frac{1}{36} \sigma ^2 t^6 \omega ^2$ [*]  
                    &    $N^{2}B^2 t^4$
                    &   $\frac{1}{36} N^{2} \sigma ^2 t^6 \omega ^2$ [* $\natural$]   
                    &   $\frac{1}{72}{\omega^2 t^4}$ [*]  
  \\
  \hline
\makecell{ bi-frequency\\ centroid}  &  $4B^{2}t^4$  [*] 
                        & $\frac{1}{18}\sigma^{2}t^{6}\omega_{s}^{2}$  [* $\natural$]   
                        &  $4 N^2 B^{2}t^4$ [* $\natural$]
                        & $\frac{1}{18} N^2 \sigma^{2}t^{6}\omega_{s}^{2}$ [*]
                        & $\frac{1}{72}\omega_s^2 t^4$ [*]\\  \hline
   separation  &     $\frac{16}{9} B^2 t^6 \omega_{r}^2$~\cite{pang2017optimal}     
               &     $\frac{8}{\pi^{4}} \sigma^{2}t^{4}$~\cite{gefen2019}  
               &      $\frac{16}{9} N^{2} B^2 t^6 \omega_{r}^2$ [* $\natural$ ]
               &     $\frac{8}{\pi^4} N^{2} \sigma ^2 t^4$ &  ${2}/{\omega_r^2}$ [*] \\
   \hline 
\end{tabular}
\end{center}
\caption{\label{tab_optimal} Summary of the QFI in the limit that $\omega t,\omega_s t,$ or $\omega_r t \ll1$, with optimal quantum control. The results denoted with an asterisk *  are new results of this work. {The symbol $\natural$ indicates extensions to the coherent signal (non-stochastic) results (Ref.~\cite{pang2017optimal}) for completeness}, and $B$ is the magnetic field strength. Details are in Supplementary Material (SM) V.
\label{tab:summary}}
\end{table*}

We consider spectral resolution of signals that are stochastic -- i.e. in each measurement the frequencies of the signal are the same, but the amplitude {suffers from noise and changes from shot-to-shot}. 
In many sensing applications (e.g.~magnetometry, gravitational wave detection, radio astronomy), signals of interest are often not deterministic. Fields such as magnetic noise from biological tissues \cite{budker2007optical} or fluctuating electromagnetic fields in materials  \cite{crooker2004spectroscopy} naturally exhibit stochastic behavior, especially in their amplitude. For instance, in biological systems or geomagnetic contexts, the field may oscillate at a known frequency (e.g.~power line frequency \cite{klanica2023magnetotelluric} or Larmor precession \cite{PhysRevResearch.2.022064}), but the amplitude may fluctuate.
Indeed, this is a common scenario in quantum sensors where state preparation may take as long as, or longer than the integration time.
Such signals have a coherence time longer than the individual experiment duration, but shorter than the time required to collect meaningful statistics. \color{black}

{For single qubits}, Gefen \textit{et al.}~\cite{gefen2019} showed that the bandwidth limit for resolving two frequencies can be surpassed by using quantum control. 
This was demonstrated experimentally in~\cite{PhysRevA.103.032419,cao2025overcoming}, showing that
quantum-enhanced frequency resolution is possible because quantum projection noise vanishes as the measured state approaches an eigenstate~\footnote{This was inspired by the work of Tsang \cite{Tsang2016} 
who showed that by using a structured measurement, one can surpass the diffraction limit in the spatial domain when estimating the spatial separation of two sources}. 
However, the use of entangled states in this context remains unexplored.
 For parameter estimation, we typically optimise the probe state for the parameter of interest.  
 {
 While the optimal probe can be found easily in noiseless unitary parameter estimation \cite{paris2009quantum}, the presence of noise generally requires a channel-dependent optimization over the input state. The complexity of this task grows exponentially over  the number of probes, often leading to a mathematically intractable problem \cite{PhysRevLett.113.250801}. }
Our approach overcomes this challenge using a new physical insight that maps frequency estimation to the task of estimating the noise parameter of a correlated qubit dephasing channel. 
We provide the first general and exact derivation of upper bounds for AC signal estimation in the stochastic regime. 
We determine the limits for resolving closely spaced frequencies and show that in certain regimes they can be saturated by certain classes of nonclassical states. Our findings apply to spin systems such as NV centres in diamonds~\cite{Barry2020Mar}, trapped-ions~\cite{Bruzewicz2019Jun}, neutral atoms~\cite{Adams2019Dec} and semiconductor qubits~\cite{burkard2023semiconductor}.

\color{black}
We show that in the low bandwidth limit, the GHZ state provides a quadratic improvement over unentangled states.  Superior performance is attained by a superposition of Dicke states, in analogy to a coherent state in a bosonic system. 
In some cases, this achieves the ultimate upper bound set by quantum information theory in the limit of large qubit number and small bandwidth.
Our new findings are summarized in Table~\ref{tab:summary} along with known results.

\section{The model and quantum parameter estimation}
We illustrate our method with a single or bi-frequency signal that acts on {a system of $N$} qubits. {Our results} can easily be extended to an arbitrary number of frequencies. 
For single-frequency signals the single-qubit Hamiltonian is
\begin{align}
\label{eq:single}
 H(t) = \Big[A\cos(\omega t) +B\sin(\omega t)\Big] \sigma_{z}, 
\end{align}
and for bi-frequency signals
\begin{align}
\label{eq:ham_double}
H(t) = \big[& A_{1}\cos\big(\omega_{1} t\big) + B_{1}\sin\big(\omega_{1} t\big)\nn
           &+ A_{2}\cos\big(\omega_{2} t\big) +  B_{2}\sin\big(\omega_{2} t\big)\big] \sigma_{z}.
\end{align}
For an $N$ qubit system, $\sigma_z$ is replaced by 
{\mbox{$ \sigma_Z =\sum_k^N \sigma_{z,k}$}}, 
where $\sigma_{z,k}$ is the Pauli Z operator for the $k$th qubit.  

{We define two parameters, the frequency centroid $~ \omega_{s} = \frac{\omega_{1}+\omega_{2}}{2}$ and separation
$\omega_{r} = \frac{\omega_{1}-\omega_{2}}{2}$,
such that $\omega_{1} = \omega_{r} + \omega_{s}$ and $\omega_{2} = \omega_{s} - \omega_{r}$.}
\noindent
If the signal is incoherent and stochastic, then every shot has a different phase and amplitude.
Following the model in Ref.~\cite{gefen2019}, we assume $A_i, B_i$ to be normally distributed, with zero mean and $\sigma^2$ variance. Essentially, this models an AC signal with randomized amplitude and phase.  

Suppose we let the probe state $\ket{\psi_0}$ evolve for time $t$, then under an effective Hamiltonian $H(t)$, the accumulated phase is
$\phi = \int_0^t dt' H(t') t'$. For one realization of 
$A_i, B_i$ the state is $\ket{\psi_{A_i,B_i}} = \exp(-i \sigma_Z \phi)\ket{\psi_0}$.
The state of the probe averaged over the random variables $A_i, B_i$ is:
\begin{align}
\rho = \int P(A_{i})P(B_{i}) |\psi_{A_{i}, B_{i}}\rangle \langle \psi_{A_{i}, B_{i}}| ~ dA_{i}dB_{i} \, ,
\end{align}
where 
$P(A_{i})$, $P(B_{i})$ are the probability distributions for $A_{i}$ and $B_{i}$.

For parameter estimation, the ultimate precision is specified by the quantum Cram\'er--Rao bound~\cite{caves,caves1} (see also~\cite{giovannetti2011advances,giovannetti2006quantum}).
For the estimation of a parameter $\theta$ encoded into a quantum state~$\hat \rho_\theta$, 
the quantum Cram\'er--Rao bound sets a lower bound on the variance $(\Delta \theta)^2 = \langle \theta^2 \rangle - \langle \theta \rangle^2$ of any unbiased estimator of $\theta$.
For unbiased estimators we have
$
 (\Delta  \theta) ^2 \geqslant  \frac{1}{\nu  J_\theta( \hat \rho_\theta)} \, ,
$
where $\nu$ is the number of copies of $\hat \rho_\theta$ used and $J_\theta$ is the quantum Fisher information (QFI) associated with the state $\hat \rho_\theta$. 

{The QFI is
$J_\theta(\rho_\theta)=\sum_{j,k:\lambda_j+\lambda_k\neq 0}
\frac{2|\bra{j}\rho'_\theta\ket{ k}|^2}{\lambda_j+\lambda_k}L$,
where $\rho'_\theta=\partial\rho_\theta/\partial\theta$,
$\lambda_j$ and $\ket{j}$ are the eigenvalues and eigenvectors of $\rho_\theta$.}

\begin{figure}[t]
\includegraphics[trim = 0cm 0.0cm 0cm 0cm, clip, width=1.0\linewidth]{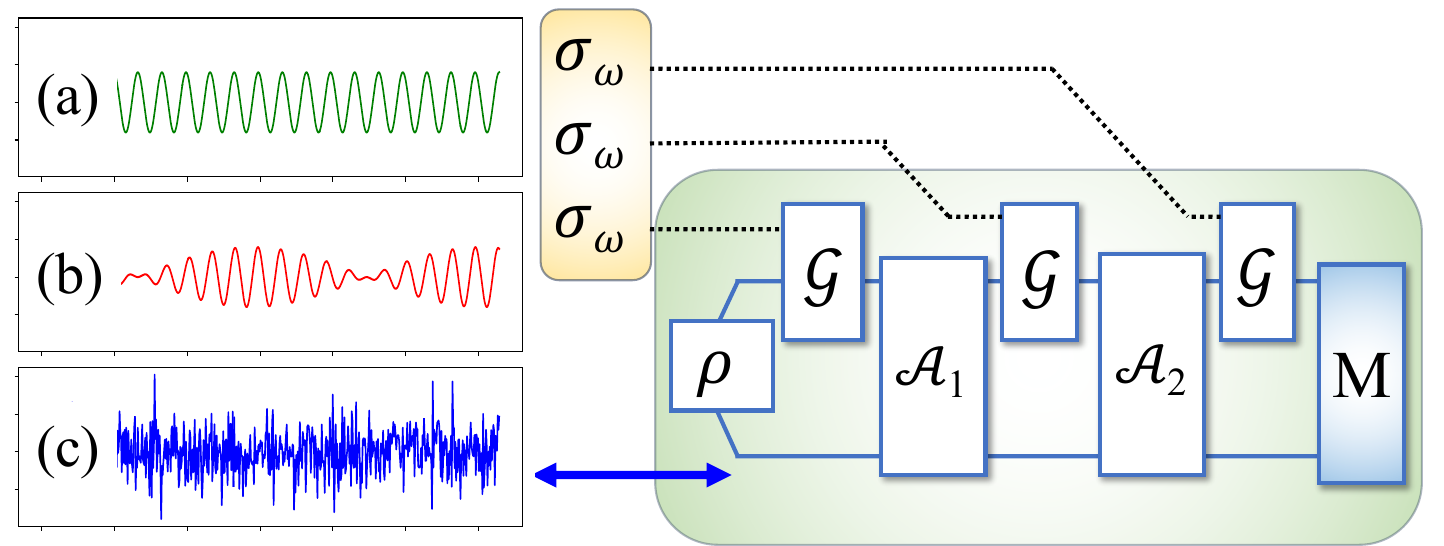} 
\caption{Left: depiction of the magnetic field signals in time, (a) A coherent single-frequency signal; (b) A coherent with bi-frequency signal; (c) An incoherent bi-frequency signal (Eq.~\eqref{eq:ham_double}). Right: a general adaptive protocol for parameter estimation. The initial state is $\rho$, the adaptive operations are $\mathcal{A}_1$ and $\mathcal{A}_2$ and the final measurement is $M$. The underlying environment states are $\sigma_\omega$, which is a mixed state that models how the stochastic signals are applied, and is not experimentally accessible.
}
\label{fig:env_state}
\end{figure}

\section{Results}

{Our first insight is that the dynamical evolution of the probes is formally represented as a global dephasing channel where all the qubits experience the same phase with some underlying probability. }
For an initial state $\rho$ consisting of $N$ qubits, this can be modeled as

\begin{align}
\label{eq:globaldephase}
\mathcal{E}[\rho_N] \rightarrow \int d\phi ~q(\phi) ~U_\phi^{\otimes N} \rho_{N} U_\phi^{\dag \otimes N},
\end{align} \color{black}
where $q(\phi)$ is the probability that the phase $\phi$ is applied,
%
and
\mbox{$U_\phi^{\otimes N} =\exp\left( -i \phi \sum_{k=1}^N  \sigma_{z,k}\right)$}.
We observe that the qubit dephasing channel can be modeled in a similar way as in Ref.~\cite{katariya2021geometric,PRXQuantum.5.020354}, where the dephasing channel is simulated by the following process:
\begin{enumerate} 
  \item A phase-shift value $\phi$ is chosen randomly
according to the probability density $q(\phi)$ in Eq.~\eqref{eq:globaldephase}.
\item  The probe state has the phase operator $U_\phi^{\otimes N}$ applied to it, and the value of $\phi$ is discarded. 
\end{enumerate}
To find an optimality bound, we define an environmental state $\sigma_\omega$
\begin{align}
\sigma_\omega := \int_{-\infty}^\infty d\phi~ q(\phi) \ket{\phi}\bra{\phi},
\end{align}
here $\{\ket{\phi} \} _\phi$ can be seen as an auxiliary orthogonal basis carrying the information about the phase-shift value $\phi$.
Then, $\mathcal{E}$ decomposes as 
\begin{align}
\mathcal{E} &= \mathcal{G}  (\rho \otimes \sigma_\omega) 
= \int_{-\infty}^\infty d\phi \, 
U_\phi^{\otimes N}
\rho
{U_\phi}^{\dag \otimes N}
\bra{\phi} \sigma_\omega \ket{\phi} \, .
\end{align}
\color{black}
%
{That is, the channel can be regarded as appending the ``environment state'' $\sigma_\omega$ to the probe $\rho$, then applying a phase shift $\phi$ conditioned on the environment state. Crucially, this decomposition lets us invoke the data-processing inequality: any adaptive protocol $\mathcal{A}$ composed of any {number of} uses of the channel and LOCC operations cannot increase the QFI. Consequently, the QFI attainable with any probe state $\rho$ is upper bounded by the QFI of the environment state itself.
Figure~\ref{fig:env_state} schematically summarises this logic chain and the resulting optimality bound~\footnote{For independent dephasing channels, see e.g.~Refs.~\cite{PhysRevLett.117.190802,matsuzaki2018quantum}; Ref.~\cite{wang2024exponential} } }.


\begin{figure}[t]
\includegraphics[trim = 1cm 0.0cm 1cm 1cm, clip, width=1.0\linewidth]{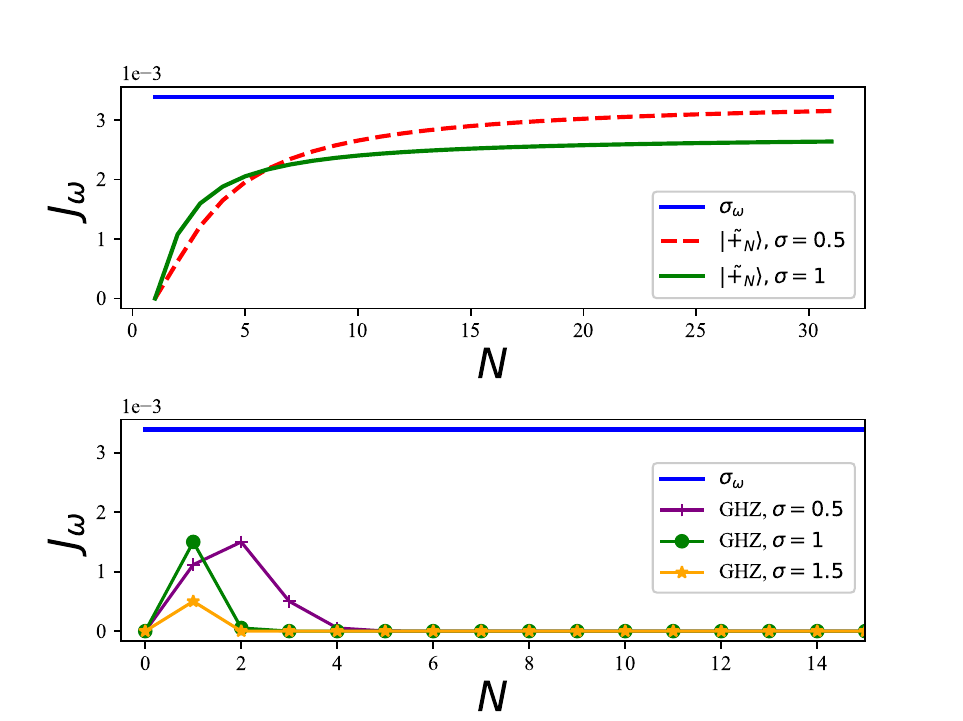} 
\caption{(Top) QFI for the spin number-state as a function of $N$. 
(Bottom) QFI for the GHZ state as a function of $N$, shown for $t = 0.7, ~\omega = 1$. The QFI of $\ket{\tilde +_N}$ is calculated numerically using the fidelity between two states (SM V,\cite{hayashi2006quantum,uhlmann1976transition}). 
}
\label{f:single_freq_results}
\end{figure}

\section{A single frequency}
First we consider estimating the frequency $\omega$ of a single-frequency stochastic signal [Eq.~(\ref{eq:single})]. After time $t$ the accumulated phase is
\begin{align}
\label{phase}
\phi_{A,B} = \frac{A}{\omega}\sin(\omega t) + \frac{B}{\omega}\big(1-\cos(\omega t)\big) .
\end{align}
$A$ and $B$ are Gaussian random variables, so $\phi_\omega$ is also a Gaussian variable.
It has zero mean, and we compute the variance to be
\begin{align}
\sigma^2_\phi 
& = 
\frac{2\sigma^2}{\omega^2} \left( 1 -\cos{(\omega t)} \right)
\, .
\end{align}
We can then compute the environment state
\begin{align}
\sigma_\omega =  \int_{-\infty}^\infty d\phi ~q(\phi) \ket{\phi}\bra{\phi},\quad
q(\phi) = \frac{1}{\sqrt{2\pi} \sigma_\phi}e^{- \frac{\phi^2}{2\sigma^2_\phi}}.
\end{align}

{Since $\sigma_\omega $ contains all the information obtainable about 
$\omega$, the quantum Fisher information for any state is bounded by the quantum Fisher information for $\sigma_\omega$, which is:}

\begin{align}
J_\omega(\sigma_\omega)  
&=\frac{2 \sqrt{2} \sin \left(\frac{\omega t }{2}\right)^5 \left(\omega t  \sin (\omega t )+2 \cos (\omega t )-2\right)^2}{\omega ^2 (1-\cos (\omega t ))^{9/2}}. 
\end{align}\color{black}
In the limit that $\omega t \ll1, J_\omega(\sigma_\omega)\approx \frac{\omega ^2 t^4}{72}$.

Ref.~\cite{PRXQuantum.5.020354} showed that for estimating bosonic dephasing channels, using a photon-number-superposition state achieves the optimality bound. 
 Here we investigate the spin number-superposition state, defined as (for details, see SM~\footnote{The SM includes Refs.~\cite{hayashi2006quantum,uhlmann1976transition} \color{black}})

\begin{align}
\ket{\tilde +_N} &= \frac{1}{\sqrt{N+1}}\sum_{n=0}^N \ket{D^N_n}, \nn
\ket{D_n^N} &= {N\choose n}^{-1/2}\sum_{ \substack{x_1,..,x_N \in {0,1}  \\ x_1+..x_N = n}}  \ket{x_1,x_2,...x_N}.
\end{align}
\color{black}
Here $\ket{D_n^N}$ is a Dicke state with excitation number $n$:  the qubit dephasing channel acts on $\ket{\tilde +_N}$ the same way a bosonic dephasing channel acts on a Fock state in that the equivalent off-diagonal terms pick up the same factor.
In the limit that $N\rightarrow \infty$, the measurement of this state converges to that of the underlying distribution $q(\phi)$,
and the number-position state saturates the precision limit \cite{PRXQuantum.5.020354}.  
The measurement is a projection onto the Fourier basis, which 
is defined for all $k \in \{0,...N\}$ as 
$\ket{u_k} = \frac{1}{\sqrt{N+1}} \sum_{n = 0}^{N+1}  e^{-2\pi i k n/(N+1)} \ket{D^N_n}$. \color{black}
We show the QFI of the spin number-position state in Fig.~\ref{f:single_freq_results} for finite $N$; we see that the upper limit is approached for certain {values of} $\sigma$'s. 
%

After the dephasing channel, the GHZ state yields only two measurement outcomes—
$\{\ket{0}^{\otimes N}\pm \ket{1}^{\otimes N}\}$—limiting the information about the channel; the off-diagonal component is suppressed by a factor $e^{-4N^2\sigma^2(...)}$ (SM, Eq 127), therefore the quality of entanglement, hence QFI decreases for large $N$.
In contrast, the spin number-superposition state retains a larger support on the Hilbert space, which allows a better estimation of the dephasing channel.
\color{black}

 We compare this to {a single} qubit state. 
 In a Ramsey experiment, the probe state of a single qubit initialised as
 $\rho_1 = \ket{+}\bra{+},~ \ket{+}=\frac{1}{\sqrt2}(\ket{0}+\ket{1})$, which evolves as
\begin{align}
\rho_1 &\rightarrow 
 \frac{1}{2}(\ket{0}\bra{0} + \gamma_1\ket{0}\bra{1} + \gamma_1{\ket{1}\bra{0}} + \ket{1}\bra{1}),
\end{align}
%
%
where
\begin{align}
\gamma_1  &= \int_{-\infty}^{\infty} \int_{-\infty}^{\infty}  dA ~dB~
e^{-2i\phi} \frac{e^ {-(A^2+B^2)/2 \sigma ^2 }}{ \left(2 \pi  \sigma^2 \right)}\nn
&=e^{-{4 \sigma ^2 (1-\cos (\omega t ))}/{\omega ^2}}.
\end{align}
The QFI for $\rho_1$ of $\omega$ is then
\begin{align}
J_\omega(\rho_1) &=
\frac{16 \sigma ^4 (\omega t  \sin (\omega t )+2 \cos (\omega t )-2)^2  \gamma_1^2  }{\omega ^6 
\left(1-\gamma_1^2\right)}.
\end{align}
In the limit that $\omega t \ll 1$, 
$J_\omega(\rho_1) \approx  \sigma ^2 t^6 \omega ^2/36$.
If a GHZ state $\rho_N =\ket{+_N}\bra{+_N}$ is used, $\ket{+_N} =\frac{1}{\sqrt2}(\ket{0}^{\otimes N} + \ket{1}^{\otimes N})$, the off-diagonal components pick up a phase factor $N$; after simplifications, the QFI computes to
\begin{align}
J_\omega(\rho_N)&=\frac{16 N^4 \sigma ^4 (\omega t  \sin (\omega t )+2 \cos (\omega t )-2)^2 \times Y}{\omega ^6 \left(1-e^{\frac{4 N^2 \sigma ^2 (\cos (\omega t )-1)}{\omega ^2}}\right)},\nn
Y &\equiv  e^{\frac{8 N^2 \sigma ^2 (\cos (\omega t )-1)}{\omega ^2}},
\end{align}
in the limit that $\omega t \ll 1$,
$J_\omega(\rho_N)\approx \frac{N^{2}}{36} \sigma ^2 t^6 \omega ^2$.

{The ratio of the $N$ to 1-qubit QFI is $J_{\omega}(\rho_N)/J_{\omega}(\rho_1)= N^{2}$, which is Heisenberg scaling.} In Fig.~\ref{f:single_freq_results}, we show the QFI for the GHZ state for different $N$, {we see that depending on $\sigma$, different $N$ is optimal.}

\section{Estimating a bi-frequency signal centroid}
For a two-frequency stochastic signal [Eq.~(\ref{eq:ham_double})], the phase accumulation is
\begin{align}
\phi =\sum_{i=1,2} 
\frac{A_i}{\omega_i} \sin{(\omega_i t)}
+ \frac{B_i}{\omega_i} \left( 1 - \cos{(\omega_i t)}\right) .
\end{align}
Calculating the variance of $\phi$ as above,
\begin{align}
\sigma^2_\phi 
& =  2\sigma^2 
\left( 
\frac{1 -\cos{(\omega_1 t)}}{\omega_1^2} 
+
\frac{1 -\cos{(\omega_2 t)}}{\omega_2^2} 
\right),
\end{align}
from which the QFI of the centroid $\omega_s = (\omega_1 +\omega_2)/2$ {is}
\begin{align}
J_{\omega_s} & = \frac{\left(\cos (\omega_r t) (\omega_s t \sin (\omega_s t)+2 \cos (\omega_s t))-2\right)^2}{2 \omega_s^2 (\cos (\omega_r t) \cos (\omega_s t)-1)^2}\\
& \approx \frac{1}{18} \sigma ^2 t^6 \omega_{s}^2.
\end{align}
This is qualitatively very similar to the single-frequency case; {using a GHZ state achieves $N^2$ scaling in the small bandwidth limit, both are as expected.}

\section{Estimating the frequency separation}
To optimally estimate the frequency separation, we apply quantum control. $\pi$ pulses are applied with frequency $\omega_s+\delta_s$ \cite{gefen2019}. 
It is important to note that this quantum control only works because the signal is coherent during each individual measurement. This leads to the effective Hamiltonian (see SM),
\begin{align}
H_\text{eff}(t) = \frac{2}{\pi}\big(& A_1\cos[(\delta_s+\delta_r)t] +  B_1 \sin[(\delta_s+\delta_r)t] + \nn
                  & A_2\cos[(\delta_s-\delta_r)t] + B_2\sin[(\delta_s-\delta_r) t] \big) \sigma_z 
\end{align}
%
Recall that $\omega_r =(\omega_1 -\omega_2)/2$,
the centroid and frequency separation become \mbox{$\delta_{s} = \frac{1}{2}(\delta_{1}+\delta_{2})$} and 
\mbox{$\delta_{r} = -\omega_{r}$}. 
Assuming $\delta_s \ll1$, the effective accumulated phase is
\begin{align}
\phi_{\omega_r} \approx \frac{ (A_2-A_1) t  \sin (\omega_r t) }{\pi ^2}.
\end{align}
%
%
We can derive the environmental state and the QFI by making the same observation about the phase,
\begin{align}
\sigma_{\phi_{\omega_r}}^2 &=
 \frac{32 \sigma ^2 t^2 \left(\omega_r^2 t^2+4 \pi ^2\right) \sin ^2\left(\frac{\omega_r t}{2}\right)}{\left(\pi  \omega_r^2 t^2-4 \pi ^3\right)^2} \\
&\approx \frac{8 \sigma ^2 t^2 \sin ^2\left(\frac{\omega_r t}{2}\right)}{\pi ^4}, \\
J_{\omega_{r}} 
&= \frac{t^2}{2\tan^2(\omega_r t/2)} \approx \frac{2}{\omega_r^2}.
\end{align}
Surprisingly, both $\sigma$ and $t$ (to first order) disappear from the optimality bound. Previously, it was found in \cite{gefen2019} that for qubit probe states, the  QFI is independent of $\omega_r$, and we have found that the QFI is in fact inversely proportional to the separation parameter, suggesting that the smaller $\omega_r$, the better it can be estimated. This agrees with the intuition from the single-qubit case~\cite{PhysRevA.103.032419} that this advantage comes from the fact that quantum projection noise vanishes as the state approaches an eigenstate. Thus, while the signal decreases, so does the noise on the measurement. It is important to note that in principle, a large number of qubits and a collective measurement would be needed to saturate this bound. 



%
For $\rho_1 = \ket{+}\bra{+}$, the off-diagonal component evolves as
\mbox{$\gamma_1 = \exp\left[-4 \sigma ^2 t^4 \omega_r^2/\pi ^4\right]$},
and the QFI is then,
\begin{equation}
J_{\omega_{r}}(\rho_1) =\frac{64 \sigma ^4 t^8 \omega_r^2}{\pi ^8 \left(e^{\frac{8 \sigma ^2 t^4 \omega_r^2}{\pi ^4}}-1\right)} \approx \frac{8\sigma^{2}t^{4}}{\pi^{4}},
\end{equation}
reproducing the main result of \cite{gefen2019}.
If the GHZ state is used, there will be a factor of $N$ in the phase term, leading to a QFI
\begin{align}
J_{\omega_{r}}(\rho_N) &=\frac{64 N^4 \sigma ^4 t^8 \text{$\omega $r}^2}{\pi ^8 \left(e^{\frac{8 N^2 \sigma ^2 t^4 \text{$\omega $r}^2}{\pi ^4}}-1\right)} 
\approx \frac{8 N^2 \sigma^{2}t^{4}}{\pi^{4}}.
\end{align}
In the small bandwidth limit, $\omega_r t \ll1$, $J_{\omega_r}(\rho_N)/J_{\omega_r}(\rho_1)= N^{2}$ which is the expected Heisenberg scaling. 

\begin{figure}[t]
\includegraphics[trim = 1cm 0.0cm 1cm 1cm, clip, width=1.0\linewidth]{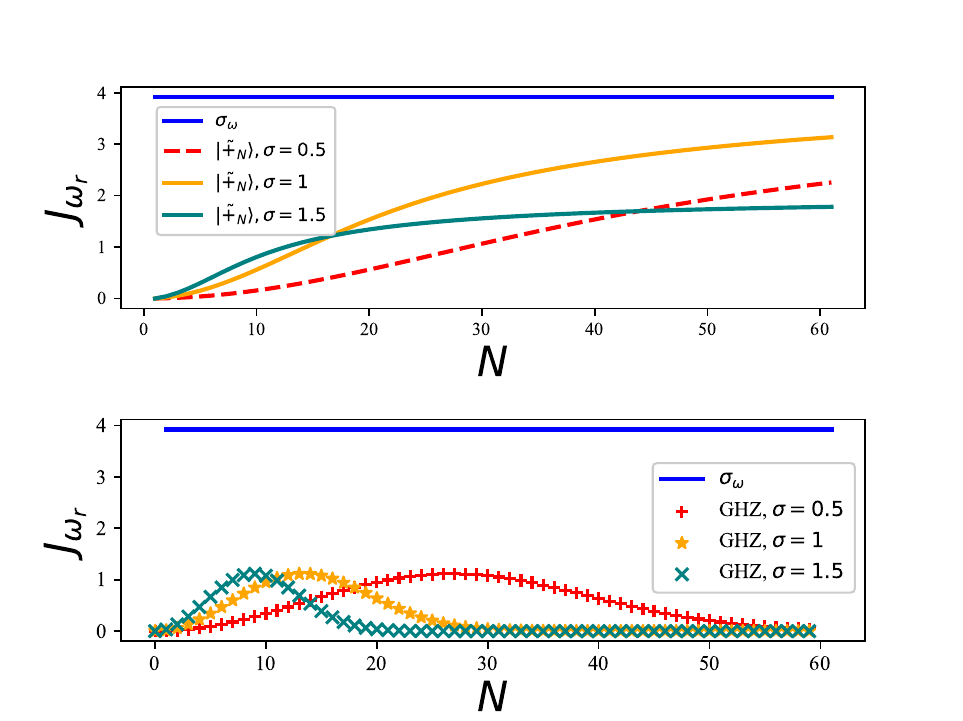} 
\caption{ (Top) QFI for the spin number-superposition state as a function of $N$, and for and (bottom) GHZ state as a function of $N$,
 shown for $t = 0.7, ~\omega_r = 0.7$. The QFI of $\ket{\tilde +_N}$ is calculated numerically using the fidelity between two states (SM V,\cite{hayashi2006quantum,uhlmann1976transition}).
 }
\label{f:separation}
\end{figure}

In Fig.~\ref{f:separation} we show the QFI of the spin number-superposition state (top) as a function of $N$, as well as different GHZ states (bottom). For the spin number-superposition state, for certain $\sigma$, the QFI plateaus ($\sigma=1$). This is likely due to a competition between phase and phase noise: for large $\sigma$, the average state decoheres quickly, and increasing the higher order components does not increase sensitivity.
This plateau does not contradict the Heisenberg scaling obtained in the low-bandwidth regime, where $\sigma t \ll1$. In this regime, using a GHZ state achieves QFI $\propto N^2$.

\color{black}

For all the parameters of interest, the optimal measurement is the projective measurement onto the Fourier basis defined above. 
For the GHZ state, optimal measurement is a projection onto the basis  $\frac{1}{\sqrt2}(\ket{0}^{\otimes N} \pm \ket{1}^{\otimes N})$.

\section{Discussions and Conclusion}
We have determined the ultimate precision bound on estimating stochastic AC signal frequencies. 
{By framing frequency estimation as a dephasing parameter estimation problem within quantum channels, we derived precise quantum Fisher information bounds for estimating frequency centroids and separations.}
We have also shown that this bound is attained by multi-qubit states in some cases, such as spin number-superposition state. 
We have also shown that entanglement is useful in the form of a $N$-qubit GHZ state, achieving $N^2$ Heisenberg scaling in the low bandwidth regime.
 Notably, despite the signal being non-unitary, the use of GHZ states in the low-bandwidth regime achieves Heisenberg scaling, underscoring the advantage of entanglement in quantum sensing.
Our findings could be easily extended to multi-frequency signals and establish a foundational framework for applying quantum metrology in AC signal processing tasks. 

{We have considered stochastic amplitudes and fixed frequencies.  The detuning parameter $\delta$ between the control field and the signal governs phase accumulation and thereby determines the effective Hamiltonian and the overall efficacy of the protocol. If the frequency is also stochastic, the control scheme loses its effectiveness. Developing alternative control strategies to handle such frequency fluctuations is a natural extension for future work.}

\begin{acknowledgments}
We thank Mark Wilde and Gavin Brennen for insightful discussions. AD is supported through Sydney Quantum Academy (SQA) primary PhD scholarship.
ZH is supported by an ARC DECRA Fellowship (DE230100144) ``Quantum-enabled super-resolution imaging''. 
CL is supported by the
European Union — Next Generation EU,
through projects PE0000023-NQSTI 
and
MUR PRIN 2022 ``QUEXO'' (CUP:D53D23002850006).
SM acknowledges support through NSF award OMA2329020.
\end{acknowledgments}


\appendix
\widetext

\section{Modelling Stochastic AC signals as a dephasing channel}

Evolution of a state under a Hamiltonian is given by

\begin{align}
\rho \rightarrow e^{-i \int H(t) dt }~\rho_0  ~ e^{i \int H(t) dt }
\end{align} 

We break down the analysis as follows
\begin{enumerate}
\item  The dynamical evolution is governed by the Hamiltonians given in Eq.(2) or (3). In the system considered here, the evolution is an accumulated phase.

\begin{align*}
\phi = \int H(t) dt.
\end{align*}

 \item Given that there are $N$ qubits in the system, for each ``run"
  of the experiment, all the qubits experience the same phase accumulation: therefore the unitary is $U^{\otimes N}$.

\item The Hamiltonian considered in this work is , each run of the experiment will have Eq.2 or 3 applied with different amplitudes. These will result in different phases being imposed. This distribution of phases we model as $q(\phi)$.


\item The overall result is that we have a collective dephasing channel described by Eq. 5.
\end{enumerate}
\color{black}

\section{ Effective Hamiltonian }
\label{sec:effective_H}
\subsection{Single frequency}
Firstly, for a single frequency signal, we reproduce in detail the derivation of the effective Hamiltonian 
\begin{align}
H_\text{eff} \approx \frac{2}{\pi}\left(A \cos[(\delta t)] + B\sin[\delta t] \right )\sigma_z.
\end{align}

The convention we use here is that for the Pauli-z operator
\begin{align}
\sigma_z= \left(
\begin{array}{cc}
 1 & 0 \\
 0 & -1 \\
\end{array}
\right),\qquad
e^{-i\phi \sigma_z} = \left(
\begin{array}{cc}
 e^{-i \phi } & 0 \\
 0 & e^{i \phi } \\
\end{array}
\right).
\end{align}

\color{black}
We start from 
\begin{align}
H(t) = [A \sin(\omega t) + B\cos(\omega t)] h(t)\sigma_z,
\end{align}
$h(t)$ is a time dependent sequence applied to manipulate the system. The Hamiltonian describes a qubit interacting with a time-varying electromagnetic field in the interaction picture. One goes into the interaction picture by applying $U = e^{-i\frac{Z}{2}\omega t}$, which absorbes the bare qubit part. The interaction part remains same as it commutes with $U$.
\begin{align}
U(t)^{\dagger}H(t)U(t) = H(t).
\end{align}

Now, the phase accumulated by the sensor depends on the driving frequency $\omega$, time $t$,
 and how frequently the $\pi$ pulses are being applied is $\tau$.

From the Hamiltonian $H(t)$ it is evident that the time dependent parts are $A \sin(\omega t) $ and $B\cos(\omega t)$.
 These two terms are responsible for the phase accumulation. 
 We know that $\cos(\omega t)$ and $\sin(\omega t)$ are the real and imaginary part of an exponential, which is a function of $\omega$ and $t$: $e^{-i\omega t}$. Hence, one can write the phase as,
\begin{align}\label{phase accumulation}
\phi = A ~\im (\Phi) + B ~ \re (\Phi), \hspace{0.3cm} \Phi = \sum\limits_{n=0}^{N-1}\int\limits_{n\tau}^{(n+1)\tau}e^{i\omega t}(-1)^{n} dt.
\end{align}
The integration part gives us the phase accumulation for a time $\tau$ and summation goes over the number of applied pulse sequences, given by $N$ here. $(-1)^{n}$ factor arrises in the equation as $e^{i m\pi} =- 1$ when $m=1,3,5,7...$ and $e^{i m\pi} =1$ when $m=0,2,4,6..$. The above equation gives us the amount of phase accumulation after application of N number of pulse sequence in interval of $\tau$.

Let us first solve the integration and the summation of the Eq.[\ref{phase accumulation}]. The integration;
\begin{eqnarray}
\int\limits_{n\tau}^{(n+1)\tau}e^{i\omega t}(-1)^{n} dt && = (-1)^{n} \frac{e^{i\omega (n+1)\tau} - e^{i\omega n\tau}}{i\omega} \nonumber \\ && = (-1)^{n}e^{i\omega n\tau} \frac{e^{i\omega \tau } -1}{i\omega}   .
\end{eqnarray}
Hence,
\begin{eqnarray}\label{phase1}
 \Phi = \sum\limits_{n=0}^{N-1}  (-1)^{n}e^{i\omega n\tau} \frac{e^{i\omega \tau } -1}{i\omega} && = \frac{e^{i\omega \tau } -1}{i\omega} \sum\limits_{n=0}^{N-1}  (-1)^{n}e^{i\omega n\tau} \nonumber \\ && = \frac{e^{i\omega \tau } -1}{i\omega} \sum\limits_{n=0}^{N-1} e^{i n \pi} e^{i\omega n\tau} \nonumber \\ && = \frac{e^{i\omega \tau } -1}{i\omega} \sum\limits_{n=0}^{N-1}  e^{i n (\omega\tau + \pi)} \nonumber \\
\end{eqnarray}
The summation $ \sum\limits_{n=0}^{N-1} e^{i n (\omega\tau + \pi)} =1+  e^{i (\omega\tau + \pi)} + e^{2 i (\omega\tau + \pi)} + e^{3 i (\omega\tau + \pi)}+ ......e^{(N-1) i (\omega\tau + \pi)}$. This is a geometric series with increment by a factor $e^{i (\omega\tau + \pi)}$. Now, by using the summation formula for the geometric series, we can write this summation as,
\begin{eqnarray}
 \sum\limits_{n=0}^{N-1} e^{i n (\omega\tau + \pi)} && = 1+  e^{i (\omega\tau + \pi)} + e^{2 i (\omega\tau + \pi)} + ....e^{(N-1) i (\omega\tau + \pi)} \nonumber \\ && =  \frac{1- e^{i N (\omega\tau + \pi)}}{1-e^{i (\omega\tau + \pi)}} \nonumber \\ && = \frac{1- e^{i N (\omega\tau + \pi)}}{1+e^{i \omega \tau}} \hspace{0.3 cm} as \hspace{0.3cm} e^{i \pi} =- 1. 
\end{eqnarray}
Then, Eq.~\eqref{phase1} becomes
\begin{align}
\Phi = \frac{1- e^{i N (\omega\tau + \pi)}}{1+e^{i \omega \tau}} \frac{e^{i\omega \tau } -1}{i\omega}.
\end{align}
Let us write down the above expression as:
\begin{align}
 1- e^{i N (\omega\tau + \pi)} = 1-e^{i \theta} \nonumber ,
\end{align}
where $\theta = N (\omega\tau + \pi)$. 
\begin{eqnarray}
e^{i \theta}-1  && =  e^{i \frac{\theta}{2}}.e^{i \frac{\theta}{2}} -1 \nonumber \\ 
                && =  \left(\cos\frac{\theta}{2} + i\sin \frac{\theta}{2}\right)\left(\cos\frac{\theta}{2} + i\sin \frac{\theta}{2}\right) -1 \nonumber \\ 
                && =  2i\sin \frac{\theta}{2}\cos\frac{\theta}{2} - \sin^{2}\frac{\theta}{2} + \cos^{2}\frac{\theta}{2} -1 \nonumber \\
                && = 2i\sin \frac{\theta}{2}\cos\frac{\theta}{2} -\sin^{2}\frac{\theta}{2} - \left(1- \cos^{2}\frac{\theta}{2}\right) \nonumber \\ 
                && = 2i\sin \frac{\theta}{2}\cos\frac{\theta}{2}  - 2\sin^{2}\frac{\theta}{2} \nonumber \\ 
                && = 2i \sin \frac{\theta}{2}\left(\cos\frac{\theta}{2}+i\sin \frac{\theta}{2}\right) \nonumber \\ 
                && =2i \sin \frac{\theta}{2}e^{i \frac{\theta}{2}}.
\end{eqnarray}
Next, we have,
\begin{eqnarray}
\frac{e^{i\omega \tau } -1}{i\omega (1+e^{i \omega \tau})} && = \frac{\cos(\omega \tau) + i\sin(\omega \tau) -1}{1+\cos(\omega \tau) + \sin(\omega \tau)(i\omega)} \nonumber \\ && = \frac{-2 \sin^{2}(\frac{\omega \tau}{2}) + 2 i\sin(\frac{\omega \tau}{2}) \cos(\frac{\omega \tau}{2})}{\big(2\cos^{2}(\frac{\omega \tau}{2}) + 2 i\sin(\frac{\omega \tau}{2}) \cos(\frac{\omega \tau}{2})\big)(i\omega)} \nonumber \\ && = \frac{\sin(\frac{\omega \tau}{2})(2i \cos(\frac{\omega \tau}{2})-2\sin(\frac{\omega \tau}{2}))}{\omega \cos(\frac{\omega \tau}{2})(2i \cos(\frac{\omega \tau}{2})-2\sin(\frac{\omega \tau}{2}))} \nonumber \\ && =\frac{\sin(\frac{\omega \tau}{2})}{\omega \cos(\frac{\omega \tau}{2})}.
\end{eqnarray}
Using these trigonometric manipulation we have,
\begin{align}
\Phi = -2ie^{i\frac{N}{2}(\omega \tau + \pi)} \sin\left(N \frac{\omega \tau}{2} + N \frac{\pi}{2}\right)  \frac{\sin(\frac{\omega\tau}{2})}{\omega \cos(\frac{\omega\tau}{2})}.
\end{align}
Now, from the above equation,
\begin{eqnarray}
\re(\Phi) 
&& = 2\sin^{2}\left(N \frac{\omega \tau}{2} + N \frac{\pi}{2}\right) \frac{\sin(\frac{\omega\tau}{2})}{\omega \cos(\frac{\omega\tau}{2})}\nn 
&&  = \big(1-\cos (\omega t + N\pi)\big)\frac{\sin(\frac{\omega\tau}{2})}{\omega \cos(\frac{\omega\tau}{2})}, 
\end{eqnarray}
and
\begin{eqnarray}
\im (\Phi) && = -2\cos\left(N \frac{\omega \tau}{2} + N \frac{\pi}{2}\right)\sin\left(N \frac{\omega \tau}{2} + N \frac{\pi}{2}\right) \frac{\sin(\frac{\omega\tau}{2})}{\omega \cos(\frac{\omega\tau}{2})} \nonumber \\ && = -\sin (\omega t + N\pi)\frac{\sin(\frac{\omega\tau}{2})}{\omega \cos(\frac{\omega\tau}{2})}, 
\end{eqnarray}
We used $ N \tau = t$. 
%
%

%
\color{black}
Now, since $\pi$ pulses are being applied, we we also need to consider the phase accumulation due to these pulse. The unitary operator for
a single $\pi$ pulse is $e^{-i\frac{\pi}{2}\sigma_{z}}$.
The $\pi$ is the control pulse one applies with a frequency $\omega + \delta$, meaning a $\pi$ pulse is applied at every $\frac{\pi}{\omega + \delta}$. Here $\delta$ is referred as the detuning. So, $\omega t = \omega N \frac{\pi}{\omega + \delta} \Longrightarrow \omega t = N\pi - \delta t$. Notice here that $\frac{\pi}{\omega} = \tau$, and $\delta $ is accumulation due to evolution in between the time of period $\tau$. Hence we can then write,
\begin{align}
\re (\Phi) = \big(1-\cos (2N\pi - \delta t)\big)\frac{\tan(\frac{\omega\tau}{2})}{\omega } = \big(1-\cos (\delta t)\big)\frac{\tan(\frac{\omega\tau}{2})}{\omega },
\end{align}
and
\begin{align}
\im(\Phi) = -\sin (2N\pi - \delta t)\frac{\tan(\frac{\omega\tau}{2})}{\omega} = \sin (\delta t)\frac{\tan(\frac{\omega\tau}{2})}{\omega}.
\end{align}
Total accumulated phase is then, 
\begin{align}\label{phase3}
\phi = A \sin (\delta t)\frac{\tan(\frac{\omega\tau}{2})}{\omega} + B \big(1-\cos (\delta t)\big)\frac{\tan(\frac{\omega\tau}{2})}{\omega }.
\end{align}
Now, it is to be observed that the above phase can be achieved by an effective Hamiltonian,

\begin{align}
\label{eq:eff_single}
H_\text{eff}(t) = \tan\left(\frac{\omega\tau}{2}\right) \frac{\delta}{\omega}\big[A\cos(\delta t) + B \sin(\delta t) \big].
\end{align}
To see the phase accumulated under the above Hamiltonian, we have
\begin{align}
 \phi = \exp\left[-i\int\limits_{0}^{t} dt ~ H_\text{eff}(t)\right].
\end{align}
The integration in the exponential results:
\begin{eqnarray}
\int\limits_{0}^{t} dt ~H_\text{eff}(t)
 && = \tan\left(\frac{\omega\tau}{2}\right) \frac{\delta}{\omega}\left[\frac{A}{\delta}\sin(\delta t) + \frac{B}{\delta} (1-\cos(\delta t) \right] \nonumber \\ 
&& = \frac{\tan(\frac{\omega\tau}{2})}{\omega}\left[A\sin(\delta t) + B (1-\cos(\delta t) \right], \nonumber \\ 
\end{eqnarray}
which is the exact phase we got in the Eq.[\ref{phase3}].\\~\\
Now, for $\delta \ll \omega$, we can write,
\begin{align}
\tan(\frac{\omega\tau}{2})\Big( \frac{\delta}{\omega}\Big) = \tan\Big(\frac{\pi}{2(1+\frac{\delta}{\omega})}\Big)\Big(\frac{\delta}{\omega}\Big).
\end{align}
We can apply l'Hospital's rule to simplify the above equation for the limit $\frac{\delta}{\omega} \rightarrow 0 $. 
Let us write $\frac{\delta}{\omega} = x$, then,
\begin{eqnarray}
\lim\limits_{x\rightarrow 0} \tan\Big(\frac{\pi}{2(1+x)}\Big)x && = \lim\limits_{x\rightarrow 0} \frac{x}{\cot\Big(\frac{\pi}{2(1+x)}\Big)} \nonumber \\ && = \lim\limits_{x\rightarrow 0} \frac{1}{-\csc^{2}\Big(\frac{\pi}{2(1+x)}\Big) \times (\frac{\pi}{2} -\frac{1}{(1+x)^{2}}) } \nonumber \\ &&  = \frac{1}{\frac{\pi}{2}} = \frac{2}{\pi}.
\end{eqnarray}
Finally, the effective Hamiltonian is: 
\begin{align} 
H_\text{eff}(t) \approx \frac{2}{\pi} \Big[A \cos(\delta t) + B \sin(\delta t) \Big]\sigma_{z}.
\end{align}
and the accumulated phase is
\begin{align}
\phi = \frac{2}{\pi}\frac{\left(A\sin(\delta t) + B(1-\cos[\delta t])\right)}{\delta}
\end{align}\

Compare this to without control:
\begin{align}
\phi = \frac{A (1-\cos (\omega t ))+B \sin (\omega t )}{\omega }.
\end{align}
\color{black}

\subsection{Two frequencies}

 Let us first begin by defining two parameters,
\begin{align}
\omega_{r} = \frac{\omega_{1}-\omega_{2}}{2}, \qquad \omega_{s} = \frac{\omega_{1}+\omega_{2}}{2},
\end{align}
such that $\omega_{1} = \omega_{r} + \omega_{s}$ and $\omega_{2} = \omega_{s} - \omega_{s}$. 
%
Now, the two-frequency Hamiltonian is
\begin{align}
\label{eq:twofreq}
H(t ) = [A_1 \sin(\omega_1 t) + B_1\cos(\omega_1 t) +A_2 \sin(\omega_2 t) + B_2\cos(\omega_2 t)  ] \sigma_z.
\end{align}

\noindent
Without additional control, the natural phase accumulation $\phi_{A,B}$ is given by,
\begin{align}
\phi = \sum\limits_{i=1,2} \frac{A_{i}}{\omega_{i}}\sin(\omega_{i}t) + \frac{B_{i}}{\omega_{i}}\big(1-\cos(\omega_{i}t)\big).
\end{align}

The above equation for two different frequencies can be expressed as,
\begin{align}
\label{onetwo}
\phi = & \frac{A_{1}}{(\omega_{r} + \omega_{s})}\sin((\omega_{r} + \omega_{s})t) + \frac{A_{2}}{(\omega_{s} - \omega_{r})}\sin((\omega_{s} - \omega_{r})t) \nn 
& + \frac{B_{1}}{(\omega_{r} + \omega_{s})}\big(1-\cos((\omega_{r} + \omega_{s})t)\big)  \nn
& + \frac{B_{2}}{(\omega_{s} - \omega_{r})}\big(1-\cos((\omega_{s} - \omega_{r})t)\big),
\end{align}
The above expression can be written as by a suitably tuning the time  $t$, such that $\omega_{s}t = 2\pi n$,
\begin{align}
\phi  \approx \frac{A_{1}}{\omega_{s}}\sin(\omega_{r}t)-\frac{A_{2}}{\omega_{s}}\sin(\omega_{r}t). 
\end{align}
The only problem with the above expression is the presence of $\omega_s$ in the denominator, which can be large. \\~\\

We now use the quantum control, where the control has frequency $(\omega_s + \delta_s)$, we define 
\begin{align}
\omega_s &= (\omega_1+\omega_2)/2\\
\delta_s &= (\delta_1 + \delta_2)\\
\delta_r &= (\omega_2-\omega_1)/2 = -\omega_r
\end{align}

%
The detunings $\delta_1$ and $\delta_2$ are such that   $\big(\omega_1 + \delta_1\big)t=\pi$ and $\big(\omega_2 + \delta_2\big)t=\pi.$ Then $\big(\omega_1 + \delta_1\big)t + \big(\omega_2 + \delta_2\big)t = 2\pi \Rightarrow = \Big[\frac{\omega_1 + \omega_2}{2} + \frac{\delta_1+\delta_2}{2}\Big]t = \pi \Rightarrow \big(\omega_s + \delta_s\big)t=\pi$. 

%
%
%
\begin{align}
H_\text{eff}(t) = \frac{2}{\pi}\big(& A_1\cos[(\delta_s+\delta_r)t] +  B_1 \sin[(\delta_s+\delta_r)t] + \nn
                  & A_2\cos[(\delta_s-\delta_r)t] + B_2\sin[(\delta_s-\delta_r) t] \big) \sigma_z 
\end{align}

The phase accumulation for $H_\text{eff}$ can be similarly written as $\phi = \int_{t'=0}^t dt' H_\text{eff}(t') $
\begin{align}
\label{eq:bi2diff}
 \phi    
 = \frac{2}{\pi}\bigg[ &\frac{1}{(\delta_r + \delta_s)}\left(A_1 \sin\big((\delta_s + \delta_r )t\big)+ B_1(1-\cos\big((\delta_r + \delta_s )t\big) \right)\nonumber\\ 
+ &\frac{1}{(\delta_s - \delta_r)}\left(A_2 \sin\big((\delta_s - \delta_r )t\big)+ B_2(1-\cos\big((\delta_s - \delta_r) t\big) \right)
\bigg].
\end{align}
%
After evolution of the effective Hamiltonian, the centroid $\omega_s$ is shifted to $\delta_s = (\delta_1+ \delta_2)/2$.\\~\\
To validate the super-resolution condition, we need to have vanishing $p$, for which we tune up $\delta_{s}t = 2\pi n$. Hence the optimal strategy becomes, $\delta_{s} t = \pm 2\pi $, or $\delta_{s} = \frac{\pm 2\pi}{t}$. With this the phase $\phi$ becomes,
\begin{align}
\phi  = \frac{2}{\pi}\bigg[&\frac{1}{(\delta_r + \delta_s)}\Big(A_1 \sin\big(\delta_r t\big)+ B_1(1-\cos\big(\delta_r t\big)\big) \Big) \nn
      +&\frac{1}{(\delta_s - \delta_r)}\Big(-A_2 \sin\big(\delta_r t\big)+ B_2(1-\cos\big( \delta_r t\big)\big) \Big)
\bigg] 
\end{align}
%
%
%
In the limit that $\delta_r \ll \delta_s$, we keep $\delta_r$ to within first order, the accumulated phase is then approximately
\begin{align}
\label{eq:phi_two_freq}
\phi \approx 
 \frac{ (A_2-A_1) t  \sin (\omega_r t) }{\pi ^2}
\end{align}

The off-diagonal elements can be computed analytically, where
\begin{align}
\int d\phi q(\phi) e^{-i 2\phi} = e^{-\frac{4 \sigma ^2 t^2 \sin ^2(\omega_r t)}{\pi ^4}},\nn
\int d\phi q(\phi) e^{-i N 2\phi} = e^{-\frac{4 N^2 \sigma ^2 t^2 \sin ^2(\omega_r t)}{\pi ^4}}.
\end{align}

\section{Optimality}
Firstly, note that under the Hamiltonians being considered, a single qubit will evolve as
\begin{align}
\rho = \int d\phi~ q(\phi) U_\phi \rho_0 U_\phi^\dag 
\end{align}
where $q(\phi)$ is the probability distribution for $\phi$ and the unitary is generated by the Hamiltonian $H$. The unitary is $U_\phi=e^{-i\phi \sigma_{z}}$.
Now, for multiple qubits, what we have is indeed a correlated (global) dephasing channel since all the qubits are in the same field. For an initial state $\rho_{N}$ consisting of $N$ qubits, this can be modelled as
\begin{align}
\rho_{N} \rightarrow \int d\phi ~q(\phi) ~U_\phi^{\otimes N} \rho_{N} U_\phi^{\dag \otimes N}.
\end{align}

The channel $\mathcal{E}$ acts on a density matrix $\rho$ as ,
\begin{align}
\mathcal{E}_N(\rho)\rightarrow \int d\phi ~q(\phi) ~\exp\left( -i \phi \sum_k^N  \sigma_{z,k}\right) \rho 
                                                    \exp \left( i \phi \sum_k^N  \sigma_{z,k}\right).
\end{align}
We can define an environmental state $\sigma_\omega$
\begin{align}
\sigma_\omega := \int_{-\pi}^\pi d\phi~ q(\phi) \ket{\phi}\bra{\phi}
\end{align}
here $\{\ket{\phi} \} _\phi$ can be seen as an orthogonal basis for the phase $\phi$.
Then, $\mathcal{E}_N$ decomposes as
\begin{align}
\mathcal{E}_N &= \mathcal{G}  (\rho \otimes \sigma_\omega) \nn
\mathcal{G} (\rho_1 \otimes \rho_2) &=\int_{-\pi}^\pi d\phi' ~  \exp \left( i \phi' \sum_k^N  \sigma_{Z,k}\right)  \rho_1 
                                                           \exp \left( i \phi' \sum_k^N  \sigma_{Z,k}\right) 
                                                 \tr[\ket{\phi'}\bra{\phi'}\rho_2]
\end{align}
The environment state $\sigma_\omega$ is appended to the input state $\rho$, while $\mathcal{G}$ measures $\sigma_\omega$, and, based on the measured phase $\phi$, applies the unitary phase operator to $\rho$.

We now proceed to calculate the environmental states.

\subsection{Single-frequency}

The phase acquired is given by the formula:
\begin{align}
\phi = \frac{A}{\omega} \sin{(\omega t)}
+ \frac{B}{\omega} \left( 1 - \cos{(\omega t)}\right) \, .
\end{align}

\subsubsection{Method 1}
Since $A$ and $B$ are Gaussian random variables, $\phi_\omega$ is a Gaussian variable too.
It has zero mean and variance
\begin{align}
\sigma^2_\phi & = 
\frac{\sigma^2}{\omega^2} \sin{(\omega t)}^2
+ \frac{\sigma^2}{\omega^2} \left( 1 - \cos{(\omega t)}\right)^2 \\
& = 
\frac{2\sigma^2}{\omega^2} \left( 1 -\cos{(\omega t)} \right)
\, .
\end{align}

Therefore, 
\begin{align}
\sigma_\omega =  \int_{-\infty}^\infty d\phi q(\phi) \ket{\phi}\bra{\phi},\qquad
q(\phi) = \frac{1}{\sqrt{2\pi \sigma^2_\phi}}e^{- \phi^2/2\sigma^2_\phi}.
\end{align}

From $\sigma_\omega$ we can compute the QFI of $\omega$

\begin{align}
J(\sigma_\omega) = \frac{2 \sqrt{2} \sin ^5\left(\frac{\omega t }{2}\right) (\omega t  \sin (\omega t )+2 \cos (\omega t )-2)^2}{\omega ^2 (1-\cos (\omega t ))^{9/2}}
\end{align}

\subsubsection{Method 2}
%
%
Method 1 is suitable for probability distributions where the parameter has zero mean. Here we present another method that will also work for non-zero mean parameters, albeit more complex.
First, we need the property of the Dirac delta function under a change of variables:
\begin{align}
    \delta( x ) 
    = \delta( f(x) ) \frac{df}{dx} \, .
\end{align}


We know the relation between the parameters $A$, $B$ and $\omega$:
\begin{align}\label{phiAB}
\phi_\omega(A,B) = \frac{A}{\omega} \sin{(\omega t)}
+ \frac{B}{\omega} \left( 1 - \cos{(\omega t)}\right) \, .
\end{align}
Making $B$ the subject, from which we obtain
\begin{align}
B_\omega(A,\phi) =
\frac{ \omega \phi  - A \sin{(\omega t)} }{ 1 - \cos{(\omega t)} }
\end{align}
and
\begin{align}
\frac{\partial B_\omega}{\partial \phi} =
\frac{ \omega  }{ 1 - \cos{(\omega t)} } \, .
\end{align}


In our calculations, we start from a probability density distribution for the parameters $A$ and $B$:
\begin{align}
    P(A,B) = (2\pi \sigma^2)^{-1}
    e^{-\frac{A^2+B^2}{2\sigma^2}} \, .
\end{align}
Combining this and Eq.~(\ref{phiAB}) we can write the probability density distribution for the parameter $\phi$:
\begin{align}
    P_\omega(\phi) = \int dA \int dB \, P(A,B) \, \delta(\phi - \phi_\omega(A,B)) \, .
\end{align}
Now we apply the property of the Dirac delta function to make a change of variable from $\phi$ to $B$.
From
\begin{align}
    \delta(\phi - \phi_\omega(A,B))
    =
    \delta(B - B_\omega(A,\phi) ) 
    \, \frac{\partial B_\omega}{\partial\phi}
\end{align}
we obtain
\begin{align}
    P_\omega(\phi) 
    & = \int dA \int dB \, P(A,B) \, 
    \delta(B - B_\omega(A,\phi) ) 
    \, \frac{\partial B_\omega}{\partial\phi} \\
& = \int dA \, P(A,B_\omega(A,\phi)) \, 
\frac{\partial B_\omega}{\partial\phi} \\
& = \frac{ (2\pi \sigma^2)^{-1} \omega }{ 1 - \cos{(\omega t)} }
\int dA \, 
e^{-\frac{A^2+B_\omega(A,\phi)^2}{2\sigma^2}} \\
& =  
\frac{  (2\pi \sigma^2)^{-1} \omega}{ 1 - \cos{(\omega t)} }
\int dA \, 
\exp{\left[
-\frac{A^2 + 
\frac{ \left( \omega \phi  - A \sin{(\omega t)} \right)^2}{ \left( 1 - \cos{(\omega t)} \right)^2}
}{2\sigma^2}
\right]} \\
& = 
\frac{ (2\pi \sigma^2)^{-1} \omega  }{ 1 - \cos{(\omega t)} }
\int dA \, 
\exp{\left[
-\frac{A^2 \left( 1 - \cos{(\omega t)} \right)^2 + 
\left( \omega \phi  - A \sin{(\omega t)} \right)^2}
{2\sigma^2 \left( 1 - \cos{(\omega t)} \right)^2 }
\right]} \\
& = 
\frac{ (2\pi \sigma^2)^{-1} \omega }{ 1 - \cos{(\omega t)} }
\int dA \, 
\exp{\left[
-\frac{
2A^2 \left( 1 - \cos{(\omega t)} \right)
+ (\omega \phi)^2
- 2 A \omega \phi \sin{(\omega t)}
}
{2\sigma^2 \left( 1 - \cos{(\omega t)} \right)^2 }
\right]} \\
& = 
\frac{ (2\pi \sigma^2)^{-1} \omega }{ 1 - \cos{(\omega t)} }
\exp{\left[
-\frac{
(\omega \phi)^2}
{2\sigma^2 \left( 1 - \cos{(\omega t)} \right)^2 }
\right]} 
\int dA \, 
\exp{\left[
-\frac{
2A^2 \left( 1 - \cos{(\omega t)} \right)
- 2 A \omega \phi \sin{(\omega t)}
}
{2\sigma^2 \left( 1 - \cos{(\omega t)} \right)^2 }
\right]} 
\, .
\label{integral}
\end{align}

Our goal is to compute the Fisher information of the continuous probability density distribution $P_\omega(\phi)$ for the estimation of $\omega$.


The integral in Eq.~(\ref{integral})
is a Gaussian integral in $A$ and can be computed exactly. The result is an explicit form for $P_\omega(\phi)$ that depends only on $\omega$, $t$, and $\sigma.$
We obtain
\begin{align}
J & =
\int dA \, 
\exp{\left[
-\frac{
2A^2 \left( 1 - \cos{(\omega t)} \right)
- 2 A \omega \phi \sin{(\omega t)}
}
{2\sigma^2 \left( 1 - \cos{(\omega t)} \right)^2 }
\right]}  \\
& =
\exp{\left[
\frac{
(\omega \phi)^2 \sin{(\omega t)}^2 / [2\left( 1 - \cos{(\omega t)} \right)]
}
{2\sigma^2 \left( 1 - \cos{(\omega t)} \right)^2 }
\right]}  \nonumber \\
& \phantom{=}~\times
\int dA \, 
\exp{\left[
-\frac{
2A^2 \left( 1 - \cos{(\omega t)} \right)
- 2 A \omega \phi \sin{(\omega t)}
+ (\omega \phi)^2 \sin{(\omega t)}^2 /[ 2\left( 1 - \cos{(\omega t)} \right) ]
}
{2\sigma^2 \left( 1 - \cos{(\omega t)} \right)^2 }
\right]} 
\\
& =
\exp{\left[
\frac{
(\omega \phi)^2 \sin{(\omega t)}^2 / [2\left( 1 - \cos{(\omega t)} \right)]
}
{2\sigma^2 \left( 1 - \cos{(\omega t)} \right)^2 }
\right]}  \nonumber \\
& \phantom{=}~\times
\int dA \, 
\exp{\left[
-\frac{
\left[ A \sqrt{2\left( 1 - \cos{(\omega t)} \right)}
- (\omega \phi) \sin{(\omega t)} /\sqrt{2\left( 1 - \cos{(\omega t)} \right)} \right]^2
}
{2\sigma^2 \left( 1 - \cos{(\omega t)} \right)^2 }
\right]} 
\\
& =
\exp{\left[
\frac{
(\omega \phi)^2 \sin{(\omega t)}^2
}
{4\sigma^2 \left( 1 - \cos{(\omega t)} \right)^3 }
\right]}  
\int dA \, 
\exp{\left[
-\frac{
\left[ A 
- (\omega \phi) \sin{(\omega t)} /\left[ 2\left( 1 - \cos{(\omega t)} \right) \right] \right]^2
}
{\sigma^2 \left( 1 - \cos{(\omega t)} \right) }
\right]} 
\\
& =
\exp{\left[
\frac{
(\omega \phi)^2 \sin{(\omega t)}^2
}
{4\sigma^2 \left( 1 - \cos{(\omega t)} \right)^3 }
\right]}  
\sqrt{\pi \sigma^2 \left( 1 - \cos{(\omega t)} \right)} \, .
\end{align}

In conclusion,
\begin{align}
    P_\omega(\phi) 
& = 
\frac{ (2\pi \sigma^2)^{-1} \omega }{ 1 - \cos{(\omega t)} }
\exp{\left[
-\frac{
(\omega \phi)^2}
{2\sigma^2 \left( 1 - \cos{(\omega t)} \right)^2 }
\right]} 
\exp{\left[
\frac{
(\omega \phi)^2 \sin{(\omega t)}^2
}
{4\sigma^2 \left( 1 - \cos{(\omega t)} \right)^3 }
\right]}  
\sqrt{\pi \sigma^2 \left( 1 - \cos{(\omega t)} \right)} \\
& = 
\frac{  \omega }{ \sqrt{ 4\pi \sigma^2 \left( 1 - \cos{(\omega t)} \right) } }
\exp{\left[
- \frac{
 (\omega \phi)^2  }
{4\sigma^2 \left( 1 - \cos{(\omega t)} \right) }
\right]} \, .
\end{align}

So, we can write 
\begin{align}
\sigma_{\omega} :&= \int_{-\infty}^\infty d\phi~ \frac{  \omega }{ \sqrt{ 4\pi \sigma^2 \left( 1 - \cos{(\omega t)} \right) } }
\exp{\left[
- \frac{
 (\omega \phi)^2  }
{4\sigma^2 \left( 1 - \cos{(\omega t)} \right) }
\right]} \, \ket{\phi}\bra{\phi} 
\end{align}

Since this is the environmental state, it contains all the information obtainable about $\omega$; this gives the same result as method 1.
\begin{align}
J_\omega &= \int d \phi~ P_\omega(\phi)\left(
 \frac{\partial }{\partial \omega} \log[P_\omega(\phi)]\right)^2 \\
&=\frac{2 \sqrt{2} (\omega t  \sin (\omega t )+2 \cos (\omega t )-2)^2}{\omega ^2 (1-\cos (\omega t ))^{9/2} \left| \csc \left(\frac{\omega t }{2}\right)\right| ^5}\nonumber\\
& \approx \frac{t^4 \omega ^2}{72} + \mathcal{O}(\omega^4)
\end{align}

\color{black}

\subsection{Frequency separation}

\subsubsection{Method 1}
\begin{align}
\label{eq:freq_sep}
\phi  = \frac{2}{\pi}\bigg[&\frac{1}{(\delta_r + \delta_s)}\Big(A_1 \sin\big(\delta_r t\big)+ B_1(1-\cos\big(\delta_r t\big)\big) \Big) 
      +\frac{1}{(\delta_s - \delta_r)}\Big(-A_2 \sin\big(\delta_r t\big)+ B_2(1-\cos\big( \delta_r t\big)\big) \Big)
\bigg] 
\end{align}

Repeat the calculation above, the variance of $\eqref{eq:freq_sep}$ is
\begin{align}
\sigma_\phi^2 \approx
 \frac{8 \sigma ^2 t^2 \sin ^2\left(\frac{\omega_r t}{2}\right)}{\pi ^4}=
\frac{8 \sigma ^2 t^2 \sin ^2\left(\frac{\omega_r t}{2}\right)}{\pi ^4}
\end{align}

\begin{align}
J_{\omega_r} = \frac{1}{2} t^2 \frac{1}{\tan^2(\omega_r t/2)}
\end{align}

\subsubsection{Method 2}

Let us now follow a similar calculation in the case of differences in frequencies. 
The phase in this case reads as;
\begin{align}\label{eq:phidiff}
\phi_{\omega_{r}}(A_{i},B_{i}) \approx \frac{A_{1}}{\omega_{s}}\sin(\omega_{r}t)-\frac{A_{2}}{\omega_{s}}\sin(\omega_{r}t). 
\end{align}
We  can make one of $A_1$ or $A_2$ as subject as,
\begin{align}
A_{1\omega_{r}}(A_{2},\phi)&=A_{2}+\frac{\phi\omega_{s}}{\sin(\omega_{r}t)}.     
\end{align}
Now,
\begin{align}
\frac{\partial A_{1}}{\partial\phi}=\frac{\omega_{s}}{\sin(\omega_{r}t)}     
\end{align}

We start from a probability density distribution for the parameters $A_1$ and $A_2$:
\begin{align}
    P(A_1,A_2) = (2\pi \sigma^2)^{-1}
    e^{-\frac{A_1^2+A_2^2}{2\sigma^2}} \, .
\end{align}
Combining this and Eq.~(\ref{phidiff}) we can write the probability density distribution for the parameter $\phi$:
\begin{align}
    P_{\omega_{r}}(\phi) = \int dA_1 \int dA_2 \, P(A_1,A_2) \, \delta(\phi - \phi_{\omega_{r}}(A_1,A_2)) \, .
\end{align}
In the next step, we apply the properties of the Dirac delta function for a change of variable:
\begin{align}
\delta(\phi - \phi_{\omega_{r}}(A_1, A_2)) = \delta(A_1 - A_{1\omega_{r}}(A_1, \phi))\frac{\partial A_{1\omega_{r}}}{\partial \phi}.
\end{align}
Combining this with the probability distribution for the $\phi$, we have
\begin{align}
P_{\omega_{r}}(\phi)& =  \int dA_{2} \int dA_1 P(A_1, A_2) \delta(A_1 - A_{1\omega_{r}}(A_2, \phi))\frac{\partial A_{1\omega_{r}}}{\partial \phi}\\
& =  \int dA_2 P(A_2, A_{1\omega_{r}}(A_2, \phi)) \frac{\partial A_{1\omega_{r}}}{\partial \phi}\\
&= \frac{ (2\pi \sigma^2)^{-1} \omega_{s} }{ \sin(\omega_{r}t) }
\int dA_2 \, 
e^{-\frac{A_{2}^2+(A_{1_{\omega_{r}}}(A_2,\phi))^2}{2\sigma^2}}\\
& = \frac{ (2\pi \sigma^2)^{-1} \omega_{s} }{ \sin(\omega_{r}t) }
\int dA_{2} e^{-\frac{A_2^2 + \big(A_2 + \frac{\phi \omega_{s}}{\sin(\omega_{r}t)}\big)^{2}}{2\sigma^{2}}}\\
& = \frac{ (2\pi \sigma^2)^{-1} \omega_{s} }{ \sin(\omega_{r}t) }
\int dA_{2} \exp\Big[-\frac{2A^{2}_{2}\sin^{2}(\omega_{r}t) +2A_{2}\sin(\omega_{r}t)\phi \omega_{s} + \phi^{2} \omega_{s}^{2}} {2\sigma^{2}\sin^{2}(\omega_{r}t)}\Big]\\
& = \frac{ (2\pi \sigma^2)^{-1} \omega_{s} }{ \sin(\omega_{r}t)} \exp \Big[-\frac{\phi^{2}\omega_{s}^{2}}{2\sigma^{2}\sin^{2}(\omega_{r}t)}\Big]
\int dA_{2} \exp\Big[-\frac{2A^{2}_{2}\sin^{2}(\omega_{r}t) +2A_{2}\sin(\omega_{r}t)\phi \omega_{s}} {2\sigma^{2}\sin^{2}(\omega_{r}t)}\Big]\\
& = \frac{1}  {{\sqrt{4\pi \sigma^{2}} }} \frac{\omega_{s}}{\sin(\omega_{r}t)}\exp\Big[- \frac{\phi^{2}\omega_{s}^{2}\csc^{2}(\omega_r t)}{4\sigma^{2}}\Big]
\end{align}

So, we can write
\begin{align}
\sigma_{\omega} :&= \int_{-\infty}^\infty d\phi~ \frac{1}  {{\sqrt{4\pi \sigma^{2}} }} \frac{\omega_{s}}{\sin(\omega_{r}t)}\exp\Big[- \frac{\phi^{2}\omega_{s}^{2}\csc^{2}(\omega_r t)}{4\sigma^{2}}\Big]
\ket{\phi}\bra{\phi} 
\end{align}

Since this is the environmental state, it contains all the information obtainable about $\omega_{r}$

\begin{align}
J_{\omega_{r}} &= \int d\phi~ P_{\omega_{r}}(\phi)\left(
 \frac{\partial }{\partial \omega_{r}} \log[P_{\omega_{r}}(\phi)]\right)^2 \\
&= \frac{2 t^2 \cos ^2(\omega_{r} t)}{\sin ^2(\omega_{r}t )}
\end{align}
The above expression when $\omega_r t<<1$, gives us,
\begin{align}
J_{\omega_{r}}=\frac{2}{\omega_r^{2}}
\end{align}

\subsection{Bi-frequency centroid}

For two frequencies, the phase accumulation is
\begin{align}
\phi = \frac{A_1}{\omega_1} \sin{(\omega_1 t)}
+ \frac{B_1}{\omega_1} \left( 1 - \cos{(\omega_1 t)}\right) 
+ \frac{A_2}{\omega_2} \sin{(\omega_2 t)}
+ \frac{B_2}{\omega_2} \left( 1 - \cos{(\omega_2 t)}\right)
\, .
\end{align}
Also, in this case, $\phi$ is a Gaussian random variable with zero mean and variance
\begin{align}
\sigma^2_\phi 
& = 
2\sigma^2 
\left( 
\frac{1 -\cos{(\omega_1 t)}}{\omega_1^2} 
+
\frac{1 -\cos{(\omega_2 t)}}{\omega_2^2} 
\right)
\, .
\end{align}

The above formula is enough to compute the Fisher information matrix for the estimation of the two frequencies.
(The, with a change of variables one can obtain the Fisher matrix for the central and relative frequencies.)


If the frequency separation $\omega_r = (\omega_1 - \omega_2)/2$ is small enough, we can use an approximation:
\begin{align}
    \frac{1}{\omega_1^2}   & \simeq  \frac{1}{\omega_s^2}\left(1-\frac{2 \omega_r}{\omega_s} \right) , \\
    \frac{1}{\omega_2^2}    & \simeq    \frac{1}{\omega_s^2}\left(1+\frac{2 \omega_r}{\omega_s} \right) \, ,
\end{align}
where $\omega_2 = (\omega_1 + \omega_2)/2$.
This yields
\begin{align}
\sigma^2_\phi  & \simeq
\frac{2\sigma^2}{\omega_s^2} 
\left[ ( 1 -\cos{(\omega_1 t)}) \left( 1 - \frac{2\omega_r}{\omega_s}\right) 
+
(1 -\cos{(\omega_2 t)}) \left( 1 + \frac{2\omega_r}{\omega_s}\right) 
\right] \\
& \simeq
\frac{2\sigma^2}{\omega_2^2} 
\left[ 
2 -\cos{(\omega_1 t)}
-\cos{(\omega_2 t)}
+
\left( \cos{(\omega_1 t)}- \cos{(\omega_2 t)} \right)
\frac{2\omega_r}{\omega_s}
\right]
\, .
\end{align}

Finally, using
\begin{align}
    \cos{(\omega_1 t)}
    & = \cos{(\omega_s t)} \cos{(\omega_r t)} 
    - \sin{(\omega_s t)} \sin{(\omega_r t)} \, , \\
    \cos{(\omega_2 t)} & = \cos{(\omega_s t)} \cos{(\omega_r t)} 
    + \sin{(\omega_s t)} \sin{(\omega_r t)} \, ,
\end{align}
we obtain
\begin{align}
\sigma^2_\phi 
& \simeq
\frac{4\sigma^2}{\omega_s^2} 
\left[ 
1 
- 1\cos{(\omega_s t)} \cos{(\omega_r t)}
-
 \sin{(\omega_s t)} \sin{(\omega_r t)}
\frac{2\omega_r}{\omega_s}
\right]
\, .
\end{align}

We may also further simplify this expression by neglecting the term proportional to 
$\omega_r/\omega_s$. This yields an expression very similar to the one-signal case:
\begin{align}
\label{eq:pdfcentroid}
\sigma^2_\phi 
& \simeq
\frac{4\sigma^2}{\omega_s^2} 
\left[ 
1 
- \cos{(\omega_s t)} \cos{(\omega_r t)}
\right]
\, ,
\end{align}
where the double frequency causes beats, as we should expect! (Note that the factor is $4$ instead of $2$ simply because we have now two signals and the total intensity is doubled.
We can fix this by replacing $\sigma^2$ with $\sigma^2/2$.)
From Eq.~\eqref{eq:pdfcentroid} we can compute the QFI of $\omega_s$
\begin{align}
J_{\omega_s} = \frac{(\cos (\omega_r t) (\omega_s t \sin (\omega_s t)+2 \cos (\omega_s t))-2)^2}{2 \omega_s^2 (\cos (\omega_r t) \cos (\omega_s t)-1)^2}
\end{align}
In the limit that $\omega_r \rightarrow 0, \omega_s t\ll1$,
\begin{align}
J_{\omega_s} \approx \frac{1}{72} \omega_s^2 t^4
\end{align}

%
\section{Spin number-position states}
\label{sec:coh}

In a bosonic system, $\ket{n}$ is the Fock state with  $n$ excitations.
%
%
Then, an $N-$level number-superposition state is defined as
\begin{align}
\ket{+_N}  = \frac{1}{\sqrt {N+1}}\sum_{n=0}^{N}\ket{n}.
\end{align}

\noindent 
If we apply the bosonic phase operator where $\hat n$ is the number operator,
\begin{align}
e^{-i \hat n \phi} \ket{+_N}
&=  \frac{1}{\sqrt {N+1}}\sum_{n=0}^{N}  e^{-i \hat n \phi} \ket{n} \nn
%
\end{align}

Now, suppose we have a Fermionic system and suppose the total number of qubits is $d$ and we have $n$ excitations. Let the lower (higher) energy system in the qubit be $\ket{0}$ ($\ket{1}$). 
We need at least $n$ qubits to hold the excitations. One of these states could be
the Dicke state, which is given by an equal superposition of all possible permutations 
\begin{align}
\ket{D_n^N} = {N\choose k}^{-1/2}\sum_{ \substack{x_1,..,x_N \in {0,1}  \\ x_1+..x_N = n}}  \ket{x_1,x_2,...x_N}
\end{align}
\textcolor{blue}{Intuitively, $\ket{D_n^N}$ is one of the Fermionic analogues of $\ket{n}$.}
 Take $\ket{D_n^N}$, we have
\begin{align}
\left( -i \phi' \sum_k^N  \sigma_{z,k}\right) \ket{D_n^N} = e^{-i n\phi'} \ket{D_n^N}
\end{align}
And the qubit dephasing channel
\begin{align}
\mathcal{E}[\ket{D_k^N}\bra{D_k^N}] = \gamma_k \ket{D_k^N}\bra{D_k^N} 
= \ket{D_k^N}\bra{D_k^N}  \int d\phi' q(\phi) e^{-i 2 k \phi'}
\end{align}

Now, define a numper-superposition-like state in the basis $\ket{D_n^N}  $, which has mean excitation (spin up) approximately $(N-1)/2$,
\begin{align}
\ket{\tilde +_N} &= \frac{1}{\sqrt{N+1}}\sum_{n=0}^d \ket{D^N_n} \nn
\ket{\tilde +_N}\bra{\tilde +_N} & =\frac{1}{N+1}\sum_{n,m = 0}^N \ket{D_n^N}\bra{D_m^N}
\end{align}

$\ket{\tilde +_N}$ is the analogue for the bosonic state 
$\ket{+_N}$.
\color{black}

In the limit that \textcolor{blue}{$N\rightarrow \infty$}, Ref.~\cite{PRXQuantum.5.020354} showed that the measurement outcome converges to that of the underlying distribution $q(\phi)$, and that the coherent state is optimal.
Now, passing the state through the channel,
\begin{align}
\rho_\omega =\mathcal{E}[\ket{\tilde +_N}\bra{\tilde +_N}] = \frac{1}{N+1}
\sum_{n,m = 0}^N \gamma_{n,m}  \ket{D_n^N}\bra{D_m^N}.
\end{align}

The off-diagonal components $\gamma_{n,m}$ can be calculated analytically as per the sections above, by performing the integration
\begin{align}
\gamma_{n,m} = \int d\phi q(\phi) e^{2i (n-m) \phi}
\end{align} 

Then, we can calculate the QFI numerically via the fidelity \cite[Theorem~6.3]{hayashi2006quantum},
\begin{align}
J_\omega = \lim_{\delta\omega\rightarrow 0} \frac{8 (1-F(\rho_\omega,\rho_{\omega+\delta \omega}))}{\delta\omega^2},
\end{align}
where the fidelity between two states $\hat\rho_1$ and $\hat\rho_2$~\cite{uhlmann1976transition} is
\begin{align}
F = \tr\sqrt{\left [ \sqrt{\rho_1}\rho_2\sqrt{\rho_1}  \right]}.
\end{align}

\color{black}


\section{QFI for specific states}
A single qubit evolves as
\begin{align}
\rho = \int d\phi~ q(\phi) U_\phi \rho_0 U_\phi^\dag ,
\end{align}
where $q(\phi)$ is the probability distribution for $\phi$ and the unitary is generated by the Hamiltonian $H$, which can be written as $U=e^{-i\phi \sigma_{z}}$.
\begin{align}
\rho_0 &\rightarrow 
 \frac{1}{2}(\ket{0}\bra{0} + \gamma_1\ket{0}\bra{1} + \gamma_1\ket{0}\bra{1} + \ket{1}\bra{1}).
\end{align}
%
%

\subsection{Singe-frequency centroid}
This integration can be performed analytically,
\begin{align}
\gamma_1 =\int q(\phi) e^{-2 i \phi} &= \int_{-\infty}^{\infty} \int_{-\infty}^{\infty}  dA ~dB~
e^{-2i\phi} \frac{\exp \left(-\frac{A^2}{2 \sigma ^2}\right)}{ \left(\sqrt{2 \pi } \sigma \right)}\frac{\exp \left(-\frac{B^2}{2 \sigma ^2}\right)}{ \left(\sqrt{2 \pi } \sigma \right)} \\
&=\exp\left[-\frac{4 \sigma ^2 (1-\cos (\omega t ))}{\omega ^2}\right]
\end{align}
The density matrix after the evolution is
\begin{align}
\label{eq:n_1_centroid}
\rho =\left(
\begin{array}{cc}
 \frac{1}{2} & \frac{1}{2} e^{-\frac{4 \sigma ^2 (1-\cos (\omega t ))}{\omega ^2}} \\
 \frac{1}{2} e^{-\frac{4 \sigma ^2 (1-\cos (\omega t ))}{\omega ^2}} & \frac{1}{2} \\
\end{array}
\right).
\end{align}
%
The QFI for $\omega$ in Eq.~\eqref{eq:n_1_centroid} is 
\begin{align}
J_\omega(\rho_1) =
\frac{16 \sigma ^4 e^{-\frac{8 \sigma ^2 (1-\cos (\omega t ))}{\omega ^2}} (\omega t  \sin (\omega t )+2 \cos (\omega t )-2)^2}{\omega ^6 
\left(1-e^{-\frac{8 \sigma ^2 (1-\cos (\omega t ))}{\omega ^2}}\right)}
\end{align}
The expression in the above equation can be expanded in $\omega$ as 
\begin{align}
J_\omega(\rho_1) = \frac{16 \sigma ^4 e^{-\frac{8 \sigma ^2 (1-\cos (\omega t ))}{\omega ^2}} (\omega t  \sin (\omega t )+2 \cos (\omega t )-2)^2}{\omega ^6 
\left(1-e^{-\frac{8 \sigma ^2 (1-\cos (\omega t ))}{\omega ^2}}\right)} = \frac{\sigma ^4 t^8 \omega ^2}{9 \left(e^{4 \sigma ^2 t^2}-1\right)} + ..
\end{align}
Term containing $e^{4 \sigma ^2 t^2}$ when expanded and taken only upto the first order gives $e^{4 \sigma ^2 t^2} = 1 + 4 \sigma ^2 t^2$. Inserting this above we have 
\begin{align}
J_\omega(\rho_1) \approx \frac{\sigma ^4 t^8 \omega ^2}{9 \left(e^{4 \sigma ^2 t^2}-1\right)} = \frac{\sigma ^4 t^8 \omega ^2}{9 \left(1 + 4 \sigma ^2 t^2-1\right)} = \frac{1}{36} \sigma ^2 t^6 \omega ^2.
\end{align}

Now, what we have is effectively a correlated (global) dephasing channel since all the qubits are in the same field. For an initial state $\rho_{N}$ consisting of $N$ qubits, this can be modelled as
\begin{align}
\rho_{N} \rightarrow \int d\phi ~q(\phi) ~U_\phi^{\otimes N} \rho_{N} U_\phi^{\dag \otimes N}.
\end{align}

When a GHZ state state $\rho_N$ is used,
\begin{align}
\rho_N = \frac{1}{2}(\ket{0}^{\otimes N}\bra{0}^{\otimes N} + \gamma_N\ket{0}^{\otimes N}\bra{1}^{\otimes N} + \gamma_N\ket{0}^{\otimes N}\bra{1}^{\otimes N} + \ket{1}^{\otimes N}\bra{1}^{\otimes N}),
\end{align}
the off-diagonal components pick up a factor $\gamma_N$
\begin{align}
\gamma_N =\int d\phi ~e^{-2i N \phi} q(\phi)  &=\int_{-\infty}^{\infty}  \int_{-\infty}^{\infty}dA ~dB~
e^{-2iN\phi} \frac{\exp \left(-\frac{A^2}{2 \sigma ^2}\right)}{ \left(\sqrt{2 \pi } \sigma \right)}\frac{\exp \left(-\frac{B^2}{2 \sigma ^2}\right)}{ \left(\sqrt{2 \pi } \sigma \right)} \\
&=\exp\left[-\frac{N^{2}4 \sigma ^2 (1-\cos (\omega t ))}{\omega ^2}\right].
\end{align}

And the QFI is
\begin{align}
J_\omega(\rho_N) =\frac{16 N^4 \sigma ^4 (\omega t  \sin (\omega t )+2 \cos (\omega t )-2)^2 e^{\frac{8 N^2 \sigma ^2 (\cos (\omega t )-1)}{\omega ^2}}}{\omega ^6 \left(1-e^{\frac{4 N^2 \sigma ^2 (\cos (\omega t )-1)}{\omega ^2}}\right)}.
\end{align}
In the limit that $\omega t \ll 1$,
\begin{align}
J_\omega(\rho_N)\approx \frac{N^{2}}{36} \sigma ^2 t^6 \omega ^2
\end{align}
where the ratio between the GHZ and the single qubit state is $J_\omega(\rho_N)/J_\omega(\rho_1)={N^{2}}$, which is Heisenberg scaling.

\subsection{Bi-frequency centroid}

For the estimation of the centroid, we start with a phase without any application of the control. The phase reads as:
\begin{align}
 \phi_{A_{i}, B_{i}} = & \frac{A_{1}}{(\omega_{r} + \omega_{s})}\sin((\omega_{r} + \omega_{s})t) + \frac{A_{2}}{(\omega_{s} - \omega_{r})}\sin((\omega_{s} - \omega_{r})t) \nonumber \\ 
 & + \frac{B_{1}}{(\omega_{r} + \omega_{s})}\big(1-\cos((\omega_{r} + \omega_{s})t)\big) \nonumber \\ 
 & + \frac{B_{2}}{(\omega_{s} - \omega_{r})}\big(1-\cos((\omega_{s} - \omega_{r})t)\big).   
\end{align}
Since $\omega_s>>\omega_r$, we can approximate the above approximation as,
\begin{align}
\phi_{A_{i}, B_{i}} = & \frac{A_{1}}{ \omega_{s}}\sin((\omega_{r} + \omega_{s})t) + \frac{A_{2}}{\omega_{s}}\sin((\omega_{s} - \omega_{r})t) \nonumber \\ 
 & + \frac{B_{1}}{\omega_{s}}\big(1-\cos((\omega_{r} + \omega_{s})t)\big) \nonumber \\ 
 & + \frac{B_{2}}{\omega_{s}}\big(1-\cos((\omega_{s} - \omega_{r})t)\big).    
\end{align}
We now have,

\begin{align}
\gamma_1 & =\int q(\phi) e^{-2 i \phi}\nn
&=\int_{-\infty}^{\infty}dA_{1}~dA_{2}~dB_{1}~dB_{2}~\frac{\exp\left(-\frac{A_{1}^{2}}{2\sigma^{2}}\right)}{\left(\sqrt{2\pi}\sigma\right)}\frac{\exp\left(-\frac{A_{2}^{2}}{2\sigma^{2}}\right)}{\left(\sqrt{2\pi}\sigma\right)}\frac{\exp\left(-\frac{B_{1}^{2}}{2\sigma^{2}}\right)}{\left(\sqrt{2\pi}\sigma\right)}\frac{\exp\left(-\frac{B_{2}^{2}}{2\sigma^{2}}\right)}{\left(\sqrt{2\pi}\sigma\right)}e^{-2i\phi} \\
&=\exp\left[-\frac{8 \sigma ^2\big(1-(\cos (\omega_{r} t ) \cos (\omega_{s} t ))\big)}{\omega_{s}^2}\right].
\end{align}    
Above expression can be approximated considering that $\omega_{r}$ is small and $\omega_r t <<1$ as:
\begin{align}
\gamma_1 =\exp\left[- \frac{8 \sigma ^2\big(1-( \cos (\omega_{s} t ))\big)}{\omega_{s}^2}\right].
\end{align}
With the above phase, the density matrix becomes,
\begin{align}
\rho = \left(
\begin{array}{cc}
 \frac{1}{2} & \frac{1}{2} e^{-\frac{8 \sigma ^2\big(1- ( \cos (\omega_{s} t ))\big)}{\omega_{s}^2}} \\
 \frac{1}{2} e^{-\frac{8 \sigma ^2\big(1-( \cos (\omega_{s} t ))\big)}{\omega_{s}^2}} & \frac{1}{2} \\
\end{array}
\right).   
\end{align}
QFI with respect to the centroid $\omega_{s}$ is given by 
\begin{align}
J_{\omega_{s}}
&=\frac{64 \sigma ^4 e^{\frac{16 \sigma ^2 (\cos (\omega_{s} t)-1)}{\omega_{s}^2}} (\omega_{s} t \sin (\omega_{s} t)+2 \cos (\omega_{s} t)-2)^2}{\omega_{s}^6 \left(1-e^{\frac{16 \sigma ^2 (\cos (\omega_{s} t)-1)}{\omega_{s}^2}}\right)}\nonumber\\
&\approx \frac{4 \sigma ^4 t^8 \omega_{s}^2}{9 \left(e^{8 \sigma ^2 t^2}-1\right)}+O\left(\omega_{s}^4\right)\nonumber\\
&\approx \frac{1}{18} \sigma ^2 t^6 \omega_{s}^2.
\end{align}
This expression of QFI looks analogous to the expression of QFI for the single frequency, which is what is expected.\\~\\
Now we want to extend this case to N qubits. For N qubits, the phase picked up by the off-diagonal terms is given as:
\begin{align}
\gamma_{N} = \exp\left[-\frac{8 \sigma ^2 N^2 \big(1-(\cos (\omega_{r} t ) \cos (\omega_{s} t ))\big)}{\omega_{s}^2}\right],    
\end{align}
and the density matrix becomes,
\begin{align}
\rho = \left(
\begin{array}{cc}
 \frac{1}{2} & \frac{1}{2} e^{-\frac{8 \sigma ^2 N^2 \big(1- ( \cos (\omega_{s} t ))\big)}{\omega_{s}^2}} \\
 \frac{1}{2} e^{-\frac{8 \sigma ^2 N^2 \big(1-( \cos (\omega_{s} t ))\big)}{\omega_{s}^2}} & \frac{1}{2} \\
\end{array}
\right).    
\end{align}
Following the same calculation as above for one qubit, the QFI then becomes,
\begin{align}
J_{\omega_{s},N} & =\frac{64 N^4 \sigma ^4 (\omega_{s}t \sin (\omega t_{s})+2 \cos (\omega_{s}t)-2)^2 e^{\frac{16 N^2 \sigma ^2 (\cos (\omega_{s} t)-1)}{\omega_{s}^2}}}{\omega_{s}^6 \left(1-e^{\frac{16 N^2 \sigma ^2 (\cos (\omega_{s} t)-1)}{\omega_{s}^2}}\right)}\nonumber\\
&\approx \frac{4 N^4 \sigma ^4 t^8 \omega_{s}^2}{9 \left(e^{8 N^2 \sigma ^2 t^2}-1\right)}+O\left(\omega_{s}^4\right)\nonumber\\
& \approx \frac{1}{18} N^2 \sigma^2 t^6 \omega_{s}^2.    
\end{align}

\subsection{Bi-frequency separation}

For a two-frequency with optimal control, reading Eq.~\eqref{eq:phi_two_freq}, we have
\begin{align}
\phi \approx \frac{ (A_1-A_2) t  \sin (-(\omega_r) t) }{\pi ^2}
\end{align}

For a single-qubit state, we can compute can compute the off-diagonal element,
\begin{align}
\gamma_1 = e^{-\frac{4 \sigma ^2 t^2 \sin ^2(\omega_r t)}{\pi ^4}}
\end{align}
and the QFI computes to be 
\begin{align}
J_{\omega_r} (\rho_1) = \frac{16 \sigma ^4 t^6 \sin ^2(2 t\omega_r)}
                    {\pi ^8 \left(e^{\frac{8 \sigma ^2 t^2 \sin ^2(t\omega_r)}{\pi ^4}}-1\right)}
\approx \frac{8 \sigma ^2 t^4}{\pi ^4}
\end{align}
in the limit that $\omega_r t \ll 1 $.

Repeating the analysis for a GHZ state, we have
\begin{align}
J_{\omega_r} (\rho_N)  = 
\frac{16 N^4 \sigma ^4 t^6 \sin ^2(2 t\omega_r)}{\pi ^8 \left(e^{\frac{8 n^2 \sigma ^2 t^2 \sin ^2(t\omega_r)}{\pi ^4}}-1\right)} \approx \frac{8 N^2 \sigma ^2 t^4}{\pi ^4}
\end{align}

We can compare this with the phase accumulation without control,
%
the phase in this is
\begin{align}\label{phidiff}
\phi \approx \frac{A_{1}}{\omega_{s}}\sin(\omega_{r}t)-\frac{A_{2}}{\omega_{s}}\sin(\omega_{r}t). 
\end{align}
Performing the integration,
\begin{align}
\gamma_1 = e^{-\frac{4 \sigma ^2 \sin ^2(\omega_r t)}{\omega _s^2}}
\end{align}

\begin{align}
J_{\omega_r} = \frac{16 \sigma ^4 t^2 \sin ^2(2 \omega_r t)}{\omega_s^4 \left(e^{\frac{8 \sigma ^2 \sin ^2(\omega_r t)}{\omega_s^2}}-1\right)} 
\approx \frac{8 \sigma ^2 t^2}{\omega_s^2}
\end{align}
which is sub-optimal because $\omega_s$ could be large.


\section{Optimal QFI for coherent magnetic field}
\label{sec:coh}
We derive the QFI for GHZ states for sensing coherent magnetic fields using the procedure presented by Pang. et. al. \cite{pang2017optimal}.
\subsection*{Single frequency}
The optimal quantum fisher information (QFI) is given by the following equation,
\begin{align}\label{QFI}
J_{g}^{(Q)} \leq \Bigg[\int_{0}^{T}\Big( \mu_{max}(t)-\mu_{min}(t)\Big) dt \Bigg]^{2},
\end{align}
where $\mu_{max}$ and $\mu_{min}$ are simultaneously the maximum and minimum eigenvalues of $\partial_{g} H_{g}(t)$ and $g$ is the parameter which we want to estimate. 
The control Hamiltonian which helps to achieve the maximum QFI,
\begin{equation}\label{control}
H_{c}(t) = \sum_{k} f_{k}(t)|\psi_{k}(t)\rangle \langle \psi_{k}(t)| - H_{g}(t) + i\sum_{k}|\partial_{t}\psi_{k}(t)\rangle \langle \psi_{k}(t)|.
\end{equation}
Where $f(k)$ is a time-dependent control sequence. For a single frequency, the Hamiltonian is a qubit interacting with a uniform rotating magnetic field $\bf{B}(t) = B\Big(\cos(\omega t) \hat{x} + \sin(\omega t) \hat{z})\Big)$:
\begin{align}
H(t) = -{\bf \mu . B} = -B\Big(\cos(\omega t) \sigma_{x} + \sin(\omega t)\sigma_{z} \Big).
\end{align}
We want to estimate the frequency $\omega$ of the magnetic field. So, we are interested in $\partial_{\omega} H(t) = Bt\Big(\sin(\omega t) \sigma_{x} -\cos(\omega t)\sigma_{z}\Big)$.
The eigenvalues of $\partial_{\omega} H(t)$ Hamiltonian are $\pm Bt$ and the corresponding eigenvectors are 
\begin{align}
|\psi_{+}\rangle =& \sin\left(\frac{\omega t}{2}
\right)|0\rangle + \cos\left(\frac{\omega t}{2}\right)|1\rangle\\
|\psi_{-}\rangle =& \cos\left(\frac{\omega t}{2}\right)|0\rangle - \sin \left(\frac{\omega t}{2}\right)|1\rangle.
\end{align}
Now, we need to find the control Hamiltonian using Eq.[\ref{control}]. To do that we can simply set $f_{k}(t) = 0$, then the equation becomes,
\begin{equation}
H_{c}(t)= - H_{g}(t) + i\sum_{k}|\partial_{t}\psi_{k}(t)\rangle \langle \psi_{k}(t)|
\end{equation}
First let us calculate $ i\sum_{k}|\partial_{t}\psi_{k}(t)\rangle \langle \psi_{k}(t)|$.
\begin{align}
i\sum_{k}|\partial_{t}\psi_{k}(t)\rangle \langle \psi_{k}(t)| = & \partial_{t}|\psi_{+}(t)\rangle \langle \psi_{+}(t)| + \partial_{t}\psi_{-}(t)\rangle \langle \psi_{-}(t)|\\
=&i\Bigg( -\frac{\omega}{2}\sin^{2}\left(\frac{\omega t}{2}\right)|1\rangle \langle 0| + \frac{\omega}{2}\sin^{2}\left(\frac{\omega t}{2}\right)|1\rangle \langle 1| - \frac{\omega}{2}\cos^{2}\left(\frac{\omega t}{2}\right)|1\rangle \langle 0| + \frac{\omega}{2}\sin^{2}\left(\frac{\omega t}{2}\right)|0\rangle \langle 1|\Bigg)\\
=&\frac{\omega}{2}\sigma_{y}.
\end{align}
 The control Hamiltonian which will lead to the optimal QFI is given as
 \begin{equation}
 H_{c}(t) = -B\Big(\cos(\omega t) \sigma_{x} + \sin(\omega t)\sigma_{z} \Big) + \frac{\omega}{2}\sigma_{y}.
 \end{equation}
 The optimal QFI using Eq.[]\ref{QFI}] is
 \begin{equation}
 J_{\omega} = \Bigg[\int_{0}^{T}\Big( 2Bt\Big) dt \Bigg]^{2} = B^{2}T^{4}.
 \end{equation}
When there are N qubits and the magnetic field is acting on each qubit collectively such that the Hamiltonian is written as,
\begin{align}
H^{(N)}(t) = -B\Big(\cos(\omega t) \sum_{k=1}^{N}\sigma_{x,k} + \sin(\omega t)\sum_{k=1}^{N}\sigma_{z,k} \Big).
\end{align}
Where the operators $\sigma_{x,k}$ and $\sigma_{z,k}$ are the Pauli $x$ and $z$ operators for the $k$th qubit.\\~\\
The maximum and minimum eigenvalues for $\partial_{\omega}H^{(N)}(t)$ are given by $\pm N Bt$. With this, the QFI is 
\begin{align}
J_{\omega}^{(N)} =  \Bigg[\int_{0}^{T}\Big( 2NBt\Big) dt \Bigg]^{2} = N^{2}B^{2}T^{4}
\end{align}

 \subsection*{Bi-frequency separation}
 We consider two magnetic fields with two different frequencies, $\omega_{1}$ and $\omega_{2}$. We define $\omega_{s} = \frac{\omega_{1}+\omega_{2}}{2}$ and  $\omega_{r}=\frac{\omega_{1}-\omega_{2}}{2}$. We consider the Hamiltonian,
 \begin{align*}
 H(t) = & -B\Bigg(\Big[\cos(\omega_{1} t) +\cos(\omega_{2}t)\Big] \sigma_{x} + \Big[\sin(\omega_{1} t)+\sin(\omega_{2} t)\Big]\sigma_{z} \Bigg)\\
 =& -B\Bigg(\Big[\cos((\omega_{s}+\omega_{r}) t) +\cos((\omega_{s}-\omega_{r})t)\Big] \sigma_{x} + \Big[\sin((\omega_{s}+\omega_{r}) t)+\sin((\omega_{s}-\omega_{r}) t)\Big]\sigma_{z} \Bigg).
 \end{align*}. 
 We want to estimate the difference in two frequencies $\omega_r$. We calculate,
 \begin{equation}
 \partial_{\omega_{r}}H(t) = Bt\Bigg(\Big[\sin((\omega_{s}+\omega_{r}) t) -\sin((\omega_{s}-\omega_{r})t)\Big] \sigma_{x} + \Big[-\cos((\omega_{s}+\omega_{r}) t)+\cos((\omega_{s}-\omega_{r}) t)\Big]\sigma_{z} \Bigg)
 \end{equation}
Eigenvalues of the above operators are $\pm 2Bt\sin(\omega_{r}t)$ and the corresponding eigenvectors are given by,

\begin{align}
|\psi_{+}(t)\rangle=& \Big(1+\sin (\omega_{s} t)\Big)|0\rangle + \cos(\omega_{s} t) |1\rangle,\\
|\psi_{-}(t)\rangle=&\Big(1-\sin(\omega_{s} t )\Big)|0\rangle -\cos(\omega_{s} t)|1\rangle.
\end{align}
We need to find the control Hamiltonian which helps us to achieve the optimal QFI. We first set $f_{k}(t) = 0$ and then calculate, 
\begin{align}
i\sum_{k}|\partial_{t}\psi_{k}(t)\rangle \langle \psi_{k}(t)| = & |\partial_{t}\psi_{+}(t)\rangle \langle \psi_{+}(t)| + |\partial_{t}\psi_{-}(t)\rangle \langle \psi_{-}(t)|\\
=& i\omega_{s}\sin{2\omega_{s}t}\sigma_{z} + i\omega_{s}\cos(2\omega_{s} t) \sigma_{x} -i\omega_{s}\sigma_{x}.
\end{align}
Hence, effectively the control Hamiltonian would be,
\begin{equation}
H_{c}(t) = -H(t) + i\omega_{s}\sin{2\omega_{s}t}\sigma_{z} + i\omega_{s}\cos(2\omega_{s} t) \sigma_{x} -i\omega_{s}\sigma_{x}.
\end{equation}
Let us now calculate the optimal QFI, given by Eq.[\ref{QFI}] with the maximum and minimum eigenvalues are written as $\pm 2Bt\sin(\omega_{r}t) $.  
\begin{equation}
J_{\omega_r} = \Big[4B\int_{0}^{T}t \sin(\omega_{r}t) dt \Big]^{2}  \approx \frac{16}{9} B^2 T^6 \omega_{r}^2.
\end{equation}
When the magnetic field is acting on the N qubits collectively, the Hamiltonian is given as,
\begin{align}
H^{(N)}(t) =  -B\Bigg(\Big[\cos((\omega_{s}+\omega_{r}) t) +\cos((\omega_{s}-\omega_{r})t)\Big] \sum_{k=1}^{N}\sigma_{x,k} + \Big[\sin((\omega_{s}+\omega_{r}) t)+\sin((\omega_{s}-\omega_{r}) t)\Big]\sum_{k=1}^{N}\sigma_{z,k} \Bigg).
\end{align} 
The maximum and minimum eigenvalues for $\partial_{\omega_{r}} H^{(N)}(t)$ are given as $\pm 2NBt\sin(\omega_{r}t)$ and the QFI is given as
\begin{align}
J_{\omega_r}^{(N)} = \Big[4NB\int_{0}^{T}t \sin(\omega_{r}t) dt \Big]^{2}  \approx \frac{16}{9}  N^2 B^2 T^6 \omega_{r}^2.
\end{align}

\subsection*{Bi-frequency centroid}
For the centroid $\omega_s$, we have,
 \begin{equation}
 \partial_{\omega_{s}}H(t) = Bt\Bigg(\Big[\sin((\omega_{s}+\omega_{r}) t) +\sin((\omega_{s}-\omega_{r})t)\Big] \sigma_{x} + \Big[-\cos((\omega_{s}+\omega_{r}) t)-\cos((\omega_{s}-\omega_{r}) t)\Big]\sigma_{z} \Bigg).
 \end{equation}
The eigenvalues are $\pm 2Bt\cos(\omega_{r}t)$. 
Eigenvalues of the above operators are $\pm 2Bt\sin(\omega_{r}t)$. Corresponding eigenvectors are given by,

\begin{align*}
|\psi_{+}(t)\rangle=& \Big(1-\cos (\omega_{s} t)\Big)|0\rangle + \sin(\omega_{s} t) |1\rangle\\
|\psi_{-}(t)\rangle=&\Big(1+\cos(\omega_{s} t )\Big)|0\rangle -\sin(\omega_{s} t)|1\rangle.
\end{align*}
First we set $f_{k}(t) = 0$ as before and then calculate, 
\begin{align*}
i\sum_{k}|\partial_{t}\psi_{k}(t)\rangle \langle \psi_{k}(t)| = & |\partial_{t}\psi_{+}(t)\rangle \langle \psi_{+}(t)| + |\partial_{t}\psi_{-}(t)\rangle \langle \psi_{-}(t)|\\
=& -i\omega_{s}\sin{2\omega_{s}t}\sigma_{z} - i\omega_{s}\cos(2\omega_{s} t) \sigma_{x} -i\omega_{s}\sigma_{y}.
\end{align*}
Hence, effectively the control Hamiltonian could be,
\begin{equation}
H_{c}(t) = -H(t) -i\omega_{s}\sin{2\omega_{s}t}\sigma_{z} - i\omega_{s}\cos(2\omega_{s} t) \sigma_{x} -i\omega_{s}\sigma_{y}.
\end{equation}
The QFI is given as
\begin{equation}
J_{\omega_s} = \Big[4B\int_{0}^{T}t \cos(\omega_{r}t) dt \Big]^{2}  \approx 4B^{2} T^{4}.
\end{equation}

When there are $N$ qubits, the maximum and the minimum eigenvalues are given as $\pm 2NBt\cos(\omega_{r}t)$. The QFI is estimated to be
\begin{align}
J_{\omega_{s}}^{(N)}=\Big[4NB\int_{0}^{T}t \cos(\omega_{r}t) dt \Big]^{2}  \approx 4N^2B^{2}T^{4}.
\end{align}

\section{Probe states in the presence of noise}

In this section we investigate the effect of additional dephasing and depolarising noise on single-qubit and GHZ states. 

\subsection{Single frequency}

\subsubsection{Single qubit}

With the dephsing from the signal itself, the state $|+\rangle$,
takes the form 
\[
\rho\rightarrow\frac{1}{2}(|0\rangle\langle0|+\gamma_{1}|0\rangle\langle1|+\gamma_{1}|1\rangle\langle0|+|1\rangle\langle1|),
\]
where $\gamma_{1}=\exp\left[-\frac{4\sigma^{2}(1-\cos(\omega t))}{\omega^{2}}\right].$

The effect of additional (local i.i.d) dephasing noise will further dampen the off-diagonal term in the density matrix. It is now written as,
\[
\rho'=\frac{1}{2}(|0\rangle\langle0|+\gamma'_{1}|0\rangle\langle1|+\gamma'_{1}|1\rangle\langle0|+|1\rangle\langle1|),
\]
where $\gamma'_{1}=\exp\left[-\frac{4\sigma^{2}(1-\cos(\omega t))}{\omega^{2}}-\Gamma t\right]$
and $\Gamma$ is the local dephasing rate. We expressed the dephsing
channel as,
\[
\mathcal{E}_{\text{deph}}(\rho)=(1-p)\rho+pZ\rho Z,\quad\text{where}\quad p=\frac{1-e^{-\Gamma t}}{2},
\]
with $\Gamma$ represnting the local dephsing rate with time.\\
Now, on top of this we want to add the depolarizing noise. A depolarizing
channel acting on a state is given as,
\[
\rho^{depol}=(1-\lambda)\rho+\frac{\lambda}{2}I,
\]
The final density matrix after going thorugh the depolaring channel
with depolarizing noise $\lambda$ would be,
\begin{align*}
\rho'' & =(1-\lambda)\cdot\frac{1}{2}\begin{pmatrix}1 & \gamma'_{1}\\
\gamma'_{1} & 1
\end{pmatrix}+\lambda\cdot\frac{1}{2}\begin{pmatrix}1 & 0\\
0 & 1
\end{pmatrix}\\
 & =\frac{1}{2}\begin{pmatrix}1 & (1-\lambda)\gamma'_{1}\\
(1-\lambda)\gamma'_{1} & 1
\end{pmatrix}=\frac{1}{2}\begin{pmatrix}1 & \gamma''_{1}\\
\gamma''_{1} & 1
\end{pmatrix},
\end{align*}
with $\gamma''_{1}=\exp\left[-\frac{4\sigma^{2}(1-\cos(\omega t))}{\omega^{2}}-\Gamma t\right](1-\lambda)$.
The depolarizing channel decreases the off-diagonal term with a factor
$(1-\lambda).$ 

The QFI for this density matrix is calculated to be (upto the order
$\omega^{2}$),
\[
J_{\omega}=\frac{(\lambda-1)^{2}\sigma^{4}t^{8}\omega^{2}}{9\left(e^{4\sigma^{2}t^{2}+2\Gamma t}-(\lambda-1)^{2}\right)}\approx\frac{(1-\lambda)^{2}\sigma^{4}t^{8}\omega^{2}}{9\left((2-\lambda)\lambda+4\sigma^{2}t^{2}+2\Gamma t\right)}\quad\text{for small \ensuremath{t}}.
\]
When $\Gamma=0$ and $\lambda=0,$ it returns the result we have in
our manuscript. 
\[
J_{\omega}=\frac{1}{36}\sigma^{2}t^{6}\omega^{2}.
\]

\subsubsection{GHZ state}

A GHZ state, while go through additional dephasing and depolarising
noise, would look like,
\[
\rho_{\text{GHZ}}^{(N)}(t)=\frac{1}{2}\left(|0\rangle^{\otimes N}\langle0|^{\otimes N}+|1\rangle^{\otimes N}\langle1|^{\otimes N}+\gamma_{N}(\omega)|0\rangle^{\otimes N}\langle1|^{\otimes N}+\gamma_{N}(\omega)|1\rangle^{\otimes N}\langle0|^{\otimes N}\right),
\]
with 
\[
\gamma_{N}=\exp\left[-\frac{4N^{2}\sigma^{2}(1-\cos(\omega t))}{\omega^{2}}-N\Gamma t\right](1-\lambda)^{N}.
\]
QFI can be calculated by writing the GHZ state as effective two dimensional
matrix as,
\[
\rho''_{GHZ}=\frac{1}{2}\begin{pmatrix}1 & \gamma_{N}(\omega)\\
\gamma_{N}(\omega) & 1
\end{pmatrix}.
\]
The QFI is turned out to be (upto the order $\omega^{2}$),
\[
J_{\omega}^{N}=\frac{N^{4}\sigma^{4}t^{8}\omega^{2}(1-\lambda)^{2N}}{9\left(e^{2N\left(2N\sigma^{2}t^{2}+\Gamma t\right)}-(1-\lambda)^{2N}\right)}\approx\frac{N^{4}\sigma^{4}t^{8}\omega^{2}(1-\lambda)^{2N}}{36N^{2}\sigma^{2}t^{2}-9(1-\lambda)^{2N}+18\Gamma Nt+9}\quad\text{for small \ensuremath{t}}.
\]
Which again when $\Gamma=0$ and $\lambda=0,$ we got back our results
in the manuscript.
\[
J_{\omega}^{N}=\frac{N^{2}}{36}\sigma^{2}t^{6}\omega^{2}.
\]

\section{Bi-frequency centroid}

Follwing the above patter, the follow up would be just simple extension
of the above stuff.

\subsection{Single qubit}

Phase accumlation due to the dephasing of the signal is 
\[
\gamma_{1}=\exp\left[-\frac{8\sigma^{2}\big(1-(\cos(\omega_{r}t)\cos(\omega_{s}t))\big)}{\omega_{s}^{2}}\right]\approx\exp\left[-\frac{8\sigma^{2}\big(1-(\cos(\omega_{s}t))\big)}{\omega_{s}^{2}}\right]\quad\text{considering }\omega_{r}t<<1.
\]
After dephasing and depolarizing channel, the density matrix would
be 
\[
\rho''=\left(\begin{array}{cc}
\frac{1}{2} & \frac{1}{2}e^{-\frac{8\sigma^{2}\big(1-(\cos(\omega_{s}t))\big)}{\omega_{s}^{2}}-\Gamma t}(1-\lambda)\\
\frac{1}{2}e^{-\frac{8\sigma^{2}\big(1-(\cos(\omega_{s}t))\big)}{\omega_{s}^{2}}-\Gamma t}(1-\lambda) & \frac{1}{2}
\end{array}\right).
\]
The QFI then corresponding to this would then be,
\begin{align*}
J_{\omega_{s}}=\frac{64(\lambda-1)^{2}\sigma^{4}(t\text{\ensuremath{\omega_{s}}}\sin(t\text{\ensuremath{\omega_{s}}})+2\cos(t\text{\ensuremath{\omega_{s}}})-2)^{2}}{\text{\ensuremath{\omega_{s}}}^{6}\left(e^{2\Gamma t-\frac{16\sigma^{2}(\cos(t\text{\ensuremath{\omega_{s}}})-1)}{\text{\ensuremath{\omega_{s}}}^{2}}}-(\lambda-1)^{2}\right)} & \approx\frac{4(\lambda-1)^{2}\sigma^{4}t^{8}\text{\ensuremath{\omega}s}^{2}}{9\left(e^{8\sigma^{2}t^{2}+2\Gamma t}-(\lambda-1)^{2}\right)}\\
 & \approx\frac{4(\lambda-1)^{2}\sigma^{4}t^{8}\text{\ensuremath{\omega_{s}}}^{2}}{9\left((2-\lambda)\lambda+8\sigma^{2}t^{2}+2\Gamma t\right)}\quad\text{for small \ensuremath{t}. }
\end{align*}
Again, for $\Gamma=0$ and $\lambda=0,$ the above expression returns
the expression we have in our manusript.
\[
J_{\omega_{s}}=\frac{\sigma^{2}t^{6}\text{\ensuremath{\omega_{s}}}^{2}}{18\sigma^{2}}.
\]

\subsection{GHZ state }

For the GHZ state we have the straightforward extension. In effective
two-dimensional form we have,
\[
\rho''_{GHZ}=\left(\begin{array}{cc}
\frac{1}{2} & \frac{1}{2}e^{-\frac{8\sigma^{2}N^{2}\big(1-(\cos(\omega_{s}t))\big)}{\omega_{s}^{2}}-N\Gamma t}(1-\lambda)^{N}\\
\frac{1}{2}e^{-\frac{8\sigma^{2}N^{2}\big(1-(\cos(\omega_{s}t))\big)}{\omega_{s}^{2}}-N\Gamma t}(1-\lambda)^{N} & \frac{1}{2}
\end{array}\right).
\]
We now have the QFI as,
\begin{align*}
J_{\omega_{s}} & =\frac{64N^{4}\sigma^{4}(1-\lambda)^{2N}(t\text{\ensuremath{\omega_{s}}}\sin(t\text{\ensuremath{\omega_{s}}})+2\cos(t\text{\ensuremath{\omega_{s}}})-2)^{2}}{\text{\ensuremath{\omega_{s}}}^{6}\left(e^{\Gamma(\text{Nt}+t)-\frac{16N^{2}\sigma^{2}(\cos(t\text{\ensuremath{\omega_{s}}})-1)}{\text{\ensuremath{\omega_{s}}}^{2}}}-(1-\lambda)^{2N}\right)}\approx\frac{4N^{4}\sigma^{4}t^{8}\text{\ensuremath{\omega_{s}}}^{2}(1-\lambda)^{2N}}{9\left(e^{8N^{2}\sigma^{2}t^{2}+\Gamma(\text{Nt}+t)}-(1-\lambda)^{2N}\right)},\text{up to\ensuremath{\quad}}\ensuremath{\omega_{s}^{2}} \\
 & \approx\frac{4N^{4}\sigma^{4}t^{8}\text{\ensuremath{\omega}s}^{2}(1-\lambda)^{2N}}{9\left(8N^{2}\sigma^{2}t^{2}-(1-\lambda)^{2N}+\Gamma(\text{Nt}+t)+1\right)}\quad\text{for small \ensuremath{t}. }
\end{align*}
Again, for $\Gamma=0$ and $\lambda=0,$ the above expression returns
the expression we have in our manusript.
\[
J_{\omega_{s}}=\frac{N^{2}\sigma^{2}t^{6}\text{\ensuremath{\omega_{s}}}^{2}}{18\sigma^{2}}.
\]


\subsection{Bi-frequency separation}

\subsubsection{Single qubit}

We follow the same procedure. We have 
\[
\gamma_{1}=e^{-\frac{4\sigma^{2}t^{2}\sin^{2}(\omega_{r}t)}{\pi^{4}}}.
\]
 We apply local dephasing and depolarizing channel. After this the
density matrix takes the form,
\[
\rho''=\left(\begin{array}{cc}
\frac{1}{2} & \frac{1}{2}e^{-\frac{4\sigma^{2}t^{2}\sin^{2}(\omega_{r}t)}{\pi^{4}}-\Gamma t}(1-\lambda)\\
\frac{1}{2}e^{-\frac{4\sigma^{2}t^{2}\sin^{2}(\omega_{r}t)}{\pi^{4}}-\Gamma t}(1-\lambda) & \frac{1}{2}
\end{array}\right).
\]
The QFI with respect to $\omega_{r}$ for this state is given as,
\[
J_{\omega_{r}}=\frac{16(\lambda-1)^{2}\sigma^{4}t^{6}\sin^{2}(2t\text{\ensuremath{\omega_{r}}})}{\pi^{8}\left(e^{2t\left(\Gamma+\frac{4\sigma^{2}t\sin^{2}(t\text{\ensuremath{\omega_{r}}})}{\pi^{4}}\right)}-(\lambda-1)^{2}\right)}\approx\frac{16(\lambda-1)^{2}\sigma^{4}t^{6}(2t\text{\ensuremath{\omega_{r}}})^{2}}{\pi^{8}\left(-\lambda^{2}+2\lambda+2t\left(\Gamma+\frac{4\sigma^{2}t(t\text{\ensuremath{\omega_{r}}})^{2}}{\pi^{4}}\right)\right)},\text{for}\quad\ensuremath{\omega_{r}t<<1.} 
\]
Again, for $\Gamma=0$ and $\lambda=0,$ the above expression returns
the expression we have in our manusript.
\[
J_{\omega_{r}}=\frac{8\sigma^{2}t^{4}}{\pi^{4}}.
\]

\subsubsection{GHZ state}

Same goes for GHZ state,
\[
\rho''_{GHZ}=\left(\begin{array}{cc}
\frac{1}{2} & \frac{1}{2}e^{-\frac{4N^{2}\sigma^{2}t^{2}\sin^{2}(\omega_{r}t)}{\pi^{4}}-N\Gamma t}(1-\lambda)^{N}\\
\frac{1}{2}e^{-\frac{4N^{2}\sigma^{2}t^{2}\sin^{2}(\omega_{r}t)}{\pi^{4}}-N\Gamma t}(1-\lambda)^{N} & \frac{1}{2}
\end{array}\right)
\]
 The QFI for $\omega_{r}$ can be estimated to be 
\[
J_{\omega_{r}}=\frac{16N^{4}\sigma^{4}t^{6}(1-\lambda)^{2n}\sin^{2}(2t\text{\ensuremath{\omega_{r}}})}{\pi^{8}\left(e^{2Nt\left(\Gamma+\frac{4N\sigma^{2}t\sin^{2}(t\text{\ensuremath{\omega_{r}}})}{\pi^{4}}\right)}-(1-\lambda)^{2N}\right)}\approx\frac{16(\lambda-1)^{2}N^{2}\sigma^{4}t^{6}(2t\text{\ensuremath{\omega_{r}}})^{2}}{\pi^{8}\left(1+2Nt\left(\Gamma+\frac{4N\sigma^{2}t(t\text{\ensuremath{\omega_{r}}})^{2}}{\pi^{4}}\right)-(1-\lambda)^{2N}\right)}.
\]
The above expresison is valid for $\omega_{r}t<<1.$ \\
For $\Gamma=0$ and $\lambda=0,$ the above expression returns the
expression we have in our manusript.
\[
J_{\omega_{r}}=\frac{8N^{2}\sigma^{2}t^{4}}{\pi^{4}}.
\]

These results confirm that the QFI decreases in the presence of additional local dephasing and depolarizing noise. While both single-frequency and bi-frequency centroid estimation protocols show a modest and expected reduction in QFI, the separation protocol is more sensitive, experiencing a more noticeable decline. Nevertheless, the degradation is not catastrophic, and the protocols retain useful sensitivity even under moderate noise levels.


\begin{thebibliography}{10}

\bibitem{giovannetti2006quantum}
Vittorio Giovannetti, Seth Lloyd, and Lorenzo Maccone.
\newblock Quantum metrology.
\newblock {\em Physical review letters}, 96(1):010401, 2006.

\bibitem{giovannetti2004quantum}
Vittorio Giovannetti, Seth Lloyd, and Lorenzo Maccone.
\newblock Quantum-enhanced measurements: beating the standard quantum limit.
\newblock {\em Science}, 306(5700):1330--1336, 2004.

\bibitem{giovannetti2011advances}
Vittorio Giovannetti, Seth Lloyd, and Lorenzo Maccone.
\newblock Advances in quantum metrology.
\newblock {\em Nature photonics}, 5(4):222--229, 2011.

\bibitem{RevModPhys.89.035002}
C.~L. Degen, F.~Reinhard, and P.~Cappellaro.
\newblock Quantum sensing.
\newblock {\em Rev. Mod. Phys.}, 89:035002, Jul 2017.

\bibitem{degen2008scanning}
CL~Degen.
\newblock Scanning magnetic field microscope with a diamond single-spin sensor.
\newblock {\em Applied Physics Letters}, 92(24), 2008.

\bibitem{taylor2008high}
Jacob~M Taylor, Paola Cappellaro, Lilian Childress, Liang Jiang, Dmitry Budker,
  PR~Hemmer, Amir Yacoby, Ronald Walsworth, and MD~Lukin.
\newblock High-sensitivity diamond magnetometer with nanoscale resolution.
\newblock {\em Nature Physics}, 4(10):810--816, 2008.

\bibitem{PhysRevLett.115.170801}
Tohru Tanaka, Paul Knott, Yuichiro Matsuzaki, Shane Dooley, Hiroshi Yamaguchi,
  William~J. Munro, and Shiro Saito.
\newblock Proposed robust entanglement-based magnetic field sensor beyond the
  standard quantum limit.
\newblock {\em Phys. Rev. Lett.}, 115:170801, Oct 2015.

\bibitem{PhysRevA.96.042319}
Sanah Altenburg, Micha\l{} Oszmaniec, Sabine W\"olk, and Otfried G\"uhne.
\newblock Estimation of gradients in quantum metrology.
\newblock {\em Phys. Rev. A}, 96:042319, Oct 2017.

\bibitem{bonato2016optimized}
Cristian Bonato, Machiel~S Blok, Hossein~T Dinani, Dominic~W Berry, Matthew~L
  Markham, Daniel~J Twitchen, and Ronald Hanson.
\newblock Optimized quantum sensing with a single electron spin using real-time
  adaptive measurements.
\newblock {\em Nature nanotechnology}, 11(3):247--252, 2016.

\bibitem{jones2009magnetic}
Jonathan~A Jones, Steven~D Karlen, Joseph Fitzsimons, Arzhang Ardavan, Simon~C
  Benjamin, G~Andrew~D Briggs, and John~JL Morton.
\newblock Magnetic field sensing beyond the standard quantum limit using
  10-spin noon states.
\newblock {\em science}, 324(5931):1166--1168, 2009.

\bibitem{riberi2025optimal}
Francisco Riberi and Lorenza Viola.
\newblock Optimal asymptotic precision bounds for nonlinear quantum metrology
  under collective dephasing.
\newblock {\em APL Quantum}, 2(2), 2025.

\bibitem{gefen2019}
T.~Gefen, A.~Rotem, and A.~Retzker.
\newblock Overcoming resolution limits with quantum sensing.
\newblock {\em Nature Communications}, 10(1):4992, 2019.

\bibitem{Bonizzoni2024}
Claudio Bonizzoni, Alberto Ghirri, Fabio Santanni, and Marco Affronte.
\newblock Quantum sensing of magnetic fields with molecular spins.
\newblock {\em npj Quantum Information}, 10(1):41, 2024.

\bibitem{WangPRX}
Guoqing Wang, Yi-Xiang Liu, Jennifer~M. Schloss, Scott~T. Alsid, Danielle~A.
  Braje, and Paola Cappellaro.
\newblock Sensing of arbitrary-frequency fields using a quantum mixer.
\newblock {\em Phys. Rev. X}, 12:021061, Jun 2022.

\bibitem{Iemini2024}
Fernando Iemini, Rosario Fazio, and Anna Sanpera.
\newblock Floquet time crystals as quantum sensors of ac fields.
\newblock {\em Phys. Rev. A}, 109:L050203, May 2024.

\bibitem{WOLF2021}
F.~Wolf and P.~O. Schmidt.
\newblock Quantum sensing of oscillating electric fields with trapped ions.
\newblock {\em Measurement: Sensors}, 18:100271, 2021.

\bibitem{khodorkovsky2009decoherence}
Yuri Khodorkovsky, Gershon Kurizki, and A~Vardi.
\newblock Decoherence and entanglement in a bosonic josephson junction:
  Bose-enhanced quantum zeno control of phase diffusion.
\newblock {\em Physical Review A—Atomic, Molecular, and Optical Physics},
  80(2):023609, 2009.

\bibitem{liu2010quantum}
YC~Liu, GR~Jin, and L~You.
\newblock Quantum-limited metrology in the presence of collisional dephasing.
\newblock {\em Physical Review A—Atomic, Molecular, and Optical Physics},
  82(4):045601, 2010.

\bibitem{carnio2015robust}
Edoardo~G Carnio, Andreas Buchleitner, and Manuel Gessner.
\newblock Robust asymptotic entanglement under multipartite collective
  dephasing.
\newblock {\em Physical Review Letters}, 115(1):010404, 2015.

\bibitem{dorner2012quantum}
U~Dorner.
\newblock Quantum frequency estimation with trapped ions and atoms.
\newblock {\em New Journal of Physics}, 14(4):043011, 2012.

\bibitem{gross2010nonlinear}
Christian Gross, Tilman Zibold, Eike Nicklas, Jerome Esteve, and Markus~K
  Oberthaler.
\newblock Nonlinear atom interferometer surpasses classical precision limit.
\newblock {\em Nature}, 464(7292):1165--1169, 2010.

\bibitem{pang2017optimal}
Shengshi Pang and Andrew~N Jordan.
\newblock Optimal adaptive control for quantum metrology with time-dependent
  hamiltonians.
\newblock {\em Nature communications}, 8(1):14695, 2017.

\bibitem{budker2007optical}
Dmitry Budker and Michael Romalis.
\newblock Optical magnetometry.
\newblock {\em Nature physics}, 3(4):227--234, 2007.

\bibitem{crooker2004spectroscopy}
SA~Crooker, DG~Rickel, AV~Balatsky, and DL~Smith.
\newblock Spectroscopy of spontaneous spin noise as a probe of spin dynamics
  and magnetic resonance.
\newblock {\em Nature}, 431(7004):49--52, 2004.

\bibitem{klanica2023magnetotelluric}
Radek Klanica, Josef Pek, and Graham Hill.
\newblock Magnetotelluric power line noise removal using temporally varying
  sinusoidal subtraction of the grid utility frequency.
\newblock {\em Pure and Applied Geophysics}, 180(9):3303--3317, 2023.

\bibitem{PhysRevResearch.2.022064}
I.~I. Ryzhov, V.~O. Kozlov, N.~S. Kuznetsov, I.~Yu. Chestnov, A.~V. Kavokin,
  A.~Tzimis, Z.~Hatzopoulos, P.~G. Savvidis, G.~G. Kozlov, and V.~S. Zapasskii.
\newblock Spin noise signatures of the self-induced larmor precession.
\newblock {\em Phys. Rev. Res.}, 2:022064, Jun 2020.

\bibitem{PhysRevA.103.032419}
Sara~L. Mouradian, Neil Glikin, Eli Megidish, Kai-Isaak Ellers, and Hartmut
  Haeffner.
\newblock Quantum sensing of intermittent stochastic signals.
\newblock {\em Phys. Rev. A}, 103:032419, Mar 2021.

\bibitem{cao2025overcoming}
Qingyun Cao, Genko~T Genov, Yaoming Chu, Jianming Cai, Yu~Liu, Alex Retzker,
  and Fedor Jelezko.
\newblock Overcoming frequency resolution limits using a solid-state spin
  quantum sensor.
\newblock {\em arXiv preprint arXiv:2506.20416}, 2025.

\bibitem{Note1}
This was inspired by the work of Tsang \cite {Tsang2016} who showed that by
  using a structured measurement, one can surpass the diffraction limit in the
  spatial domain when estimating the spatial separation of two sources.

\bibitem{paris2009quantum}
Matteo~GA Paris.
\newblock Quantum estimation for quantum technology.
\newblock {\em International Journal of Quantum Information},
  7(supp01):125--137, 2009.

\bibitem{PhysRevLett.113.250801}
Rafal Demkowicz-Dobrza\ifmmode~\acute{n}\else \'{n}\fi{}ski and Lorenzo
  Maccone.
\newblock Using entanglement against noise in quantum metrology.
\newblock {\em Phys. Rev. Lett.}, 113:250801, Dec 2014.

\bibitem{Barry2020Mar}
John~F. Barry, Jennifer~M. Schloss, Erik Bauch, Matthew~J. Turner, Connor~A.
  Hart, Linh~M. Pham, and Ronald~L. Walsworth.
\newblock {Sensitivity optimization for NV-diamond magnetometry}.
\newblock {\em Rev. Mod. Phys.}, 92(1):015004, March 2020.

\bibitem{Bruzewicz2019Jun}
Colin~D. Bruzewicz, John Chiaverini, Robert McConnell, and Jeremy~M. Sage.
\newblock {Trapped-ion quantum computing: Progress and challenges}.
\newblock {\em Appl. Phys. Rev.}, 6(2), June 2019.

\bibitem{Adams2019Dec}
C.~S. Adams, J.~D. Pritchard, and J.~P. Shaffer.
\newblock {Rydberg atom quantum technologies}.
\newblock {\em J. Phys. B: At. Mol. Opt. Phys.}, 53(1):012002, December 2019.

\bibitem{burkard2023semiconductor}
Guido Burkard, Thaddeus~D Ladd, Andrew Pan, John~M Nichol, and Jason~R Petta.
\newblock Semiconductor spin qubits.
\newblock {\em Reviews of Modern Physics}, 95(2):025003, 2023.

\bibitem{caves}
Samuel~L. Braunstein and Carlton~M. Caves.
\newblock Statistical distance and the geometry of quantum states.
\newblock {\em Phys. Rev. Lett.}, 72:3439--3443, May 1994.

\bibitem{caves1}
IR~Afnan, R~Banerjee, Samuel~L Braunstein, I~Brevik, Carlton~M Caves,
  B~Chakraborty, Ephraim Fischbach, Lee Lindblom, GJ~Milburn, SD~Odintsov,
  et~al.
\newblock article no. 0048.
\newblock {\em Ann. Phys.}, 247:447, 1996.

\bibitem{katariya2021geometric}
Vishal Katariya and Mark~M. Wilde.
\newblock Geometric distinguishability measures limit quantum channel
  estimation and discrimination.
\newblock {\em Quantum Information Processing}, 20(2):78, 2021.

\bibitem{PRXQuantum.5.020354}
Zixin Huang, Ludovico Lami, and Mark~M. Wilde.
\newblock Exact quantum sensing limits for bosonic dephasing channels.
\newblock {\em PRX Quantum}, 5:020354, Jun 2024.

\bibitem{Note2}
For independent dephasing channels, see e.g.~Refs.~\cite
  {PhysRevLett.117.190802,matsuzaki2018quantum}; Ref.~\cite
  {wang2024exponential}.

\bibitem{hayashi2006quantum}
Masahito Hayashi.
\newblock {\em Quantum Information: An Introduction}.
\newblock Springer, 2006.

\bibitem{uhlmann1976transition}
Armin Uhlmann.
\newblock The ``transition probability'' in the state space of a *-algebra.
\newblock {\em Reports on Mathematical Physics}, 9(2):273--279, 1976.

\bibitem{Note3}
The SM includes Refs.~\cite {hayashi2006quantum,uhlmann1976transition}.

\bibitem{Tsang2016}
M.~Tsang, R.~Nair, and X.-M. Lu.
\newblock Quantum theory of superresolution for two incoherent optical point
  sources.
\newblock {\em Phys. Rev. X}, 6:031033, Aug 2016.

\bibitem{PhysRevLett.117.190802}
Cosmo Lupo and Stefano Pirandola.
\newblock Ultimate precision bound of quantum and subwavelength imaging.
\newblock {\em Phys. Rev. Lett.}, 117:190802, Nov 2016.

\bibitem{matsuzaki2018quantum}
Yuichiro Matsuzaki, Shiro Saito, and William~J Munro.
\newblock Quantum metrology at the heisenberg limit with the presence of
  independent dephasing.
\newblock {\em arXiv preprint arXiv:1809.00176}, 2018.

\bibitem{wang2024exponential}
Yu-Xin Wang, Jacob Bringewatt, Alireza Seif, Anthony~J Brady, Changhun Oh, and
  Alexey~V Gorshkov.
\newblock Exponential entanglement advantage in sensing correlated noise.
\newblock {\em arXiv preprint arXiv:2410.05878}, 2024.

\end{thebibliography}
\end{document}